\documentclass[letterpaper]{article}
\usepackage{aaai2027}
\usepackage[hyphens]{url}
\usepackage{graphicx}
\usepackage{natbib}
\usepackage{caption}
\usepackage{booktabs}
\usepackage{array}
\newenvironment{ack}{\section*{Acknowledgments}}{}

\title{Comprehension Audits to Mitigate Risks from Automated AI Research}
\author{
    Ronald J.~Bodkin\textsuperscript{\rm 1}\corresponding,
    Bahrad A.~Sokhansanj\textsuperscript{\rm 2},
    Gillian Hadfield\textsuperscript{\rm 3}
}
\affiliations{
    \textsuperscript{\rm 1}Independent\\
    \textsuperscript{\rm 2}Institute for Law and AI\\
    \textsuperscript{\rm 3}Johns Hopkins University\\
    rbodkin@gmail.com
}

\nocopyright
\begin{document}

\maketitle

\begin{abstract}

AI is already writing a majority of code for frontier AI labs. This creates a safety risk if there is insufficient human oversight. Existing work proposes minimum comprehension thresholds and unaided checks to mitigate this. To our knowledge, however, there is currently no published frontier-AI assurance regime that requires demonstrated evidence that the responsible humans understand what they are building as a precommitted condition for continuing development or usage. We propose \emph{comprehension audits}, a novel development-process assurance mechanism in which the responsible people explain R\&D contributions to auditors to demonstrate understanding. With independent administration and graded reports, they provide a gate: development of a contribution stops based on a failure to demonstrate human understanding until remediated, with escalating consequences for repeated failures. Our analysis of leading open-source AI projects finds increased output of code with reduced human review commentary rates per line of code, with far lower rates for automated fleet accounts. We advocate for labs to conduct them with embedded independent auditors.

\end{abstract}
\section{Introduction}
\label{sec:intro}

This paper addresses development-time assurance for frontier AI R\&D, with a focus on ensuring human comprehension to mitigate risk. It introduces comprehension audits, a novel auditing approach that tests human understanding of research contributions, which complements safety assessments of research artifacts.

Anthropic, OpenAI and Google report that AI is already writing 75\% or more of their new code \citep{Anthropic_2026a, Pichai_2026, Brockman_2026}. Automation of software development and other research activities may give rise to increased risk if there is insufficient oversight and accountability. There is also evidence that the volume of AI code produced in leading open source AI R\&D projects is greatly increasing and that human review is being stretched under this volume (as described in Sections~\ref{sec:background} and~\ref{sec:empirical} below).

Comprehension audits have an independent party select an important component, experiment or modification and conduct a review with the responsible contributors. Audits are graded based on their level of understanding. If the grade fails to meet a minimum threshold, the contributors pause development and any release that depends on the affected artifacts, until rectifying the failure. A pattern of repeated failures escalates to suspending the responsible unit's release authorization and requires rollback of affected releases and adoption of additional safeguards until the pattern of failure is rectified. In this way, the audits provide strong incentives to ensure adequate understanding of research output. They provide a gate against a systematic degradation of understanding and accountability.

Existing work \citep{Lin_2026} proposes minimum comprehension thresholds and unaided checks as tests for insufficient comprehension. To our knowledge, there is currently no published frontier-AI assurance regime that requires demonstrated evidence that the responsible humans understand what they are building as a precommitted condition for continuing development or updates. A sampling-based approach to audits allows detection of systematic failures in understanding with modest process overhead. This paper supplies that missing gate, specifies its operation and reports evidence of declining review attention.

\section{The Problem: Volume Overwhelms Oversight}
\label{sec:problem}

The International Dialogues on AI Safety (IDAIS) issued a consensus statement \citep{IDAIS_2024} that included ``No AI system should be able to copy or improve itself without explicit human approval and assistance.''

Bainbridge \citep{Bainbridge_1983} argued how increased automation can increase the difficulty in overseeing systems, notably by cognitive atrophy. \citet{Kosmyna_2025} showed empirical evidence that reliance on LLMs reduced recall, mental engagement, and self-reported ownership in the context of essay writing.

\citet{Shen_2026} conducted randomized experiments for how using AI affected developers learning a new programming library. They found ``that AI use impairs conceptual understanding, code reading, and debugging abilities, without delivering significant efficiency gains on average.'' They also found ``that AI-enhanced productivity is not a shortcut to competence and AI assistance should be carefully adopted into workflows to preserve skill formation -- particularly in safety-critical domains.'' Similarly, Mitchell et al. argue ``that current approaches to the development and deployment of AI agent systems do not support effective human oversight -- they contribute to its degradation'' \citep{Mitchell_2026}.

In reporting on increased automation of AI R\&D, Anthropic predicts that ``if they can't review code as quickly as Claude can generate it, human review will become the bottleneck to AI development'' \citep{Anthropic_2026a}. In this report, one Anthropic employee describes: ``there are days where everything breaks and I don't understand why and I realize I have no idea what I've been up to anymore.'' Likewise, OpenAI researcher Daniel Selsam stated ``... human researchers are losing the ability and the will to take true ownership of model-driven research. Researchers and engineers in all parts of the stack are rapidly increasing their dependence on the models even to perceive the world. I myself barely look at raw code anymore...'' \citep{Selsam_2026}. And the OpenAI researcher pseudonymously known as roon said ``...we have really no choice but to ask another astra to read/use the outputs of astra 1. this is relatively recent. i think even 1-1.5 generations ago people were reading the code'' (sic) \citep{Roon_2026}.

In framing the problem of \emph{scalable oversight} \citep{Amodei_2016, Christiano_2018}, AI researchers anticipated a problem in which AI produces output beyond human comprehension. Keeping up with the increased \emph{volume} of AI output is functionally in danger of overwhelming human oversight long before AI is able to create output that is beyond expert understanding.

If researchers are not aware of AI generated output they are less likely to identify misaligned output. This includes reward hacking \citep{METR_2025} such as incomplete implementations \citep{Greenblatt_2026}, faking failing tests \citep{Baker_2025}, destructive actions \citep{Anthropic_2026b}, circumventing system controls \citep{Wang_2025}, or evidence of in-context scheming \citep{Meinke_2024}. Researchers who do not review AI output carefully are vulnerable to incorporating accidental flaws in their contributions. These can include misunderstanding requirements or not catching misspecifications of requirements (e.g.,\ from not providing necessary context). These are long-standing risks in human R\&D efforts but require humans to know how to effectively work with AI agents (which are novel and frequently change) as well as occurring in a context of increased development output. These risks are upstream of any deployment of a model developed using automated R\&D. Flaws in AI R\&D artifacts that propagate into trained models are costly to remediate (as it necessitates retraining) and may not be detectable.

Ultimately, approval of AI-written output without comprehension is meaningless. Meaningful adherence to the IDAIS statement requires organizations to maintain human comprehension and accountability for AI-written output to comply. As frontier labs move away from human code review (Section~\ref{sec:trends}), establishing sufficient comprehension and accountability is a key challenge.

\section{Background and Related Work}
\label{sec:background}

In this section we look at the dynamics and risks of \emph{recursive self-improvement} (RSI) \citep{Bostrom_2014}, existing governance and audit approaches for AI and gaps therein.

\subsection{Trends and Risks of Automating AI R\&D}
\label{sec:trends}

Automation of AI R\&D poses a risk by accelerating capabilities, including the possibility of loss of control \citep{Bengio_2024}. Full automation allows for the possibility of RSI. Frontier models such as Mythos \citep{Anthropic_2026b} and GPT-6 Astra \citep{OpenAI_2026b} are being used internally by labs in advance of public deployment, with recent high profile incidents \citep{OpenAI_2026e, Anthropic_2026e} occurring for models that were not publicly deployed. Accordingly, risks from automated AI R\&D are not addressed by governance requirements that are gated on public deployment. Anthropic reported that as of August 2026 Claude ``leads'' 26\% of their R\&D work and that the share of work that is at ``AI collaborates'' or higher is above 90\% \citep{Anthropic_2026d}, using Epoch AI's automation scale \citep{Denain_2026}. The definition of leads is ``completing most of the task end to end from a high-level ask; human supervises, course-corrects, and approves outputs,'' which does not require understanding. Dario Amodei said ``AI has been advancing drastically faster, driven primarily by AI's growing ability to build the next generation of AI. This dynamic is called recursive self-improvement, and it is starting to happen across the industry, including at Anthropic'' \citep{Amodei_2026}.

In a recent survey among 25 AI researchers by Field, Douglas, and Krueger \citep{Field_2026}, 14 of 25 ranked automated AI R\&D as a serious or primary AI risk per Figure 3 in that survey. Researchers expressed skepticism about the ability to define effective red lines for the risk of automated AI R\&D, e.g.,\ one researcher said ``The more concrete your red line, the more decoupled it becomes from the abstract intelligence explosion risk that you're worried about.'' 

\citet{Lin_2026} propose the existence of a \emph{Cognitive Integrity Threshold}. As automation increases, human oversight declines and they posit that there exists this critical threshold for understanding below which oversight becomes empty. They propose testing whether comprehension is above this threshold by performing periodic verification with unaided comprehension checks and propose institutional corrective action as well as proactive mitigation measures. Their framework establishes a general comprehension principle but does not specify how to apply it for frontier-AI assurance nor an auditing process with binding gates. Comprehension audits operationalize that principle through organizationally independent audit assessments, categorical findings, appeal, scoped remediation and verification required for resumption.

\citet{Casserini_2026} identify the problem of ``... \emph{agentic entropy}, systemic drift that traditional code diff-based and HCXAI methods fail to capture...'' They propose novel tooling in response that is designed to improve human review and understanding of agentic code output.

Boris Cherny, the creator of Claude Code, posted ``Steps of AI Adoption'' \citep{Cherny_2026} and observed that ``Anthropic is on step 3 and pushing toward 4.'' Its step 3 guardrails are a combination of automatic reviews and controlling inputs to AI agents. Its step 3 development process is automated with people managing by exception only and none requires human-understandable documentation. Step 4 ``AI-native'' is summarized as ``The loop is fully closed and most agents are kicked off by Claude. Hundreds to thousands of agents run; you steer by intent and monitor by exception.'' There is no verification of human understanding at these steps even as agentic output exceeds the capacity for human code review.

\subsection{Measuring Automation and Oversight}
\label{sec:measuring}

Chan et al. propose metrics to track the extent of AI R\&D automation and its effects \citep{Chan_2026}, which are lagging proxies for how much AI has automated R\&D such as AI performance on R\&D evaluations, compute efficiency improvements, and staff views. Any such metrics for R\&D automation pose a difficult challenge for how to act on them: what value of metrics indicates excessive risk? This motivates comprehension audits as a specific, actionable threshold for adequacy.

They also propose \emph{oversight red-teaming}, in which organizations deliberately detect ``whether oversight systems used in production would catch an AI system that is directed to subvert the R\&D process, such as by attempting to sabotage experiments'' \citep{Chan_2026}. As described, this emphasizes tests for whether automated oversight systems would detect subversive actions, rather than the understanding of humans in the loop.

Chan et al. also propose \emph{oversight effectiveness retrospectives}, in which organizations analyze defects and the extent of human and automated review of those defects. They propose metrics for defect rate and time to discovery stratified by defect severity. This is a means to analyze and track the impact of reduced human attention in the development processes but as they acknowledge these are lagging indicators. They do not present a crisp gate for when automation is excessive.

Anthropic has reported metrics for contributor productivity \citep{Anthropic_2026a}, the role of AI in development, the ability to oversee and intervene in actions taken by AI agents for R\&D and resources used \citep{Anthropic_2026d}. OpenAI \citep{OpenAI_2026d} has reported similar metrics including contributor productivity, experiment velocity, metrics around use of coding agents by researcher and R\&D task, task success rates and required interventions by estimated task duration. The visibility into automated R\&D is valuable, but none of the metrics addresses staff understanding. The intervention rate is the closest proxy, but it does not distinguish between comprehension of generated code and iteration based solely on output or failures. Neither company has proposed thresholds at which these metrics would trigger limits. This paper includes an analysis of human review changes against evidence of AI usage for leading open source AI R\&D projects up to June 2026.

Existing approaches to risk governance for frontier AI have proposed pauses or mitigations based on conditions being reached \citep{Alaga_2023, Karnofsky_2024, Anthropic_2026c}. Thus far the conditions have been technical qualities of the output, such as dangerous-capability evaluations or evidence of misalignment; recent pacing proposals have added input proxies such as capability checkpoints \citep{Amodei_2026} and compute caps for capabilities research and a mandatory lag before a model may be used for AI R\&D \citep{Lifland_2026}. None of these proposals pace development based on requiring evidence that the responsible humans understand what they are building as a precommitted condition.

\subsection{Audits and Other Assurance Regimes}
\label{sec:assurance}

There is a rich research literature on audits for AI systems in general \citep{Raji_2020, Sandvig_2014, Metaxa_2021} and for Frontier AI in particular \citep{Mokander_2023, Brundage_2026}. There is also research on applying mature risk management to AI \citep{Schuett_2023} and on applying internal audits for AI risks \citep{Schuett_2025}. Mature model risk management includes maintaining documentation and a log of material issues encountered once a model is in operation and consistently remediating them. Alaghmandan and Streltchenko summarize key tenets of model risk management for financial institutions \citep{Alaghmandan_2025} that are expected under the Federal Reserve and OCC's supervisory guidance \citep{FederalReserve_2011, OCC_2026}. The 2026 revision of this guidance notes that ``Generative AI and agentic AI models are novel and rapidly evolving. As such, they are not within the scope of this guidance.'' Even the most mature risk management discipline has excluded generative and agentic AI from their scope.

ISO/IEC 42001:2023 (AI management systems) \citep{ISO_2023} is a certifiable standard that audits whether management processes exist but does not address the understanding of the staff involved. Anthropic \citep{Anthropic_2025} and OpenAI \citep{OpenAI_2026c} hold accredited 42001 certifications that cover model development as well as products.

There is a rich literature on third party verification and audits for AI \citep{Raji_2022, Brundage_2020} and the role of independent verification organizations to perform them (IVOs) \citep{Ball_2025, Fathom_2026, Stosz_2025}. \citet{Hadfield_2026} call for governments to regulate AI through mandates to have government-licensed private regulators (IVOs) verify compliance by AI developers.

The AI Evaluator Forum has published AEF-1 \citep{Stosz_2025}. This is a voluntary ``standard and checklist that third-party evaluators can use to demonstrate how they achieved a set of operating conditions that support a baseline level of independence, access, and transparency during an evaluation.'' It defines access conditions for the AI systems, data and documentation but does not require access to the development process, underlying software code or the people involved in creating the AI systems.

\citet{Brundage_2026} define a series of increasingly rigorous assurance levels ranging from AAL-1, limited assurance, to AAL-4, very high assurance. At level AAL-2 and above, it calls for third-party auditors to have deep access for at least months-long engagements focused on governance, risk management, security and safety processes as well as systems and artifacts. It calls for auditors to have authority to conduct staff interviews to understand how processes work in practice. To our knowledge this is the most expansive audit scope called for in the existing literature. None have proposed auditing the understanding of the people involved.

The EU AI Act, article 14, requires that high-risk systems be designed so the natural persons overseeing them can understand their capabilities and limitations, although the extent of understanding is not well specified \citep{EU_2024}. There are industry regulations that certify knowledge and people including for nuclear operators \citep{NRC_2024}, pharmaceutical manufacturers \citep{EU_2001} and financial statements \citep{USCongress_2002}. However, these regimes certify understanding of engineered systems with well-understood theoretical principles. By contrast, state-of-the-art AI systems are less understood. As \citet{Judge_2025} observe, ``the systems' behavior is an emergent property. \ldots{}its behavior is neither expressed intentionally by designers in software program code nor legible (yet) by examining the program code and its massive array of tuned parameters.''

\subsection{Safety Interventions}
\label{sec:interventions}

\citet{Amodei_2016} define scalable oversight as the problem of effectively using limited amounts of human oversight to align the training of AI models. \citet{Bowman_2022} define scalable oversight as ``the problem of supervising systems that potentially outperform us on most skills relevant to the task at hand.'' The field shows progress in limited scenarios \citep{Kenton_2024}, but there is no documented case of such scalable oversight protocols being used by frontier AI labs in production. We are not aware of any literature that addresses scaling \emph{human} oversight to large volumes of AI output, such as the code and experiments of automated AI R\&D.

Greenblatt et al. propose control protocols that are intended to be robust to untrustworthy LLMs that intentionally subvert them \citep{Greenblatt_2023}. Control techniques are being applied to the runtime execution of AI agents \citep{AISI_2025, FARAI_2026}. Korbak et al. present a structure that could be used to argue that AI control techniques are sufficient to allow deploying potentially misaligned powerful AI \citep{Korbak_2025a}. However, no such safety case has been demonstrated. Section~\ref{sec:discussion} below argues that control techniques are complementary to comprehension audits, with each reducing risks at a different stage of the R\&D cycle. Control techniques do not provide a gate for when risks from automated AI R\&D have become excessive.

\section{Comprehension Audits}
\label{sec:comprehension-audits}

Comprehension audits are designed to counteract declining human oversight in frontier AI research. Comprehension audits are novel compared to other kinds of AI audits previously proposed because the object of the audit is the collective \emph{human understanding} of the development team, not the artifacts they produce. Their contribution is the combination of organizationally independent assessment of comprehension with an appealable gate on the affected R\&D.

The auditors monitor ongoing development to identify candidate \emph{contributions} for a comprehension audit. A contribution is a meaningful unit of R\&D work for which a person or team is accountable and which changes the organization's products or beliefs. This can include completed experiments and their conclusions, a change to training recipe, integration of a new dataset, or optimization of a system. It is defined by the purpose and effect, not by the size or how it is produced. It may consist of hundreds of commits with many pull requests or hundreds of concurrent agent runs. Whoever directed the work and accepted its results is accountable for the contribution as a whole, regardless of how much of the work AI performed.

After the auditors have reviewed a contribution, they may schedule an audit meeting with the \emph{core contributors}: the human team members responsible for it. The purpose of the meeting is to verify that among the contributors there is adequate understanding, i.e.,\ that some responsible human understands all key aspects. Due to the collaborative nature of AI R\&D efforts, it is not reasonable to have a standard that a single named individual can answer for all aspects of a contribution, but it is important that collectively the humans involved can do so.

The audit is similar in nature to an oral examination in a thesis defense or to a design review. The contributors are expected to show understanding of the contribution without using AI assistance during the review. There are three tiers for depth of understanding required by the contributors:

\begin{table}[htbp]
  \caption{Tiers of required understanding for audited contributors.}
  \label{tab:tiers}
  \centering
  \small
  \begin{tabular}{>{\raggedright\arraybackslash}p{0.16\linewidth}>{\raggedright\arraybackslash}p{0.42\linewidth}>{\raggedright\arraybackslash}p{0.32\linewidth}}
    \toprule
    Tier & Required depth & Aspects covered \\
    \midrule
    Deep understanding & A thorough understanding of these aspects of the contribution & Architecture and design; implementation behavior; process \\
    \addlinespace
    Familiarity & Good familiarity and appropriate diligence and understanding & Key findings; analysis of results \\
    \addlinespace
    Limited & Contributors are not expected to explain root causes of model behavior or otherwise exceed the level of understanding of deep learning that prevails in the industry & Model internals; root causes of model behavior \\
    \bottomrule
  \end{tabular}
\end{table}

\citet{Mosier_1996} showed experimental evidence that when people perceive themselves to be accountable for decisions they increase verification efforts and reduce automation bias. Accordingly, we anticipate that the fact an organization conducts comprehension audits will incentivize R\&D staff to improve understanding by increasing organizational accountability. This is a secondary benefit we anticipate, in addition to the primary function of the audits of providing a way to detect and respond to the danger from insufficient comprehension.

In the next section, we detail the comprehension audit process and how it integrates into the organization and overall governance processes.

\section{Design and Operationalization}
\label{sec:design}

In this section we describe comprehension audits in more detail. The overall process is illustrated in Figure~\ref{fig:lifecycle} below. Details of operationalization including accountability and thresholds should be included in the frontier lab's safety framework.

Comprehension audits are designed to be performed by external auditors. This requires labs to embed auditors with meaningful internal access as OpenAI and Anthropic committed to \citep{Amodei_2026, OpenAI_2026f}. More generally, an AAL-2 assurance level would provide the necessary access and a months-long duration sufficient to conduct them. Labs could also adopt comprehension audits internally with audits conducted by an internal oversight function or a formal internal audit team.

An audit review board is responsible for oversight of auditor performance. This might be a regulator if mandated, an accrediting body for external auditors otherwise, or an internal oversight team.

The standards for levels of comprehension should not vary regardless of the assurance level or the organization conducting the audits: adequate human comprehension is an important standard to maintain.

\begin{figure}[htbp]
  \centering
  \includegraphics[width=\linewidth]{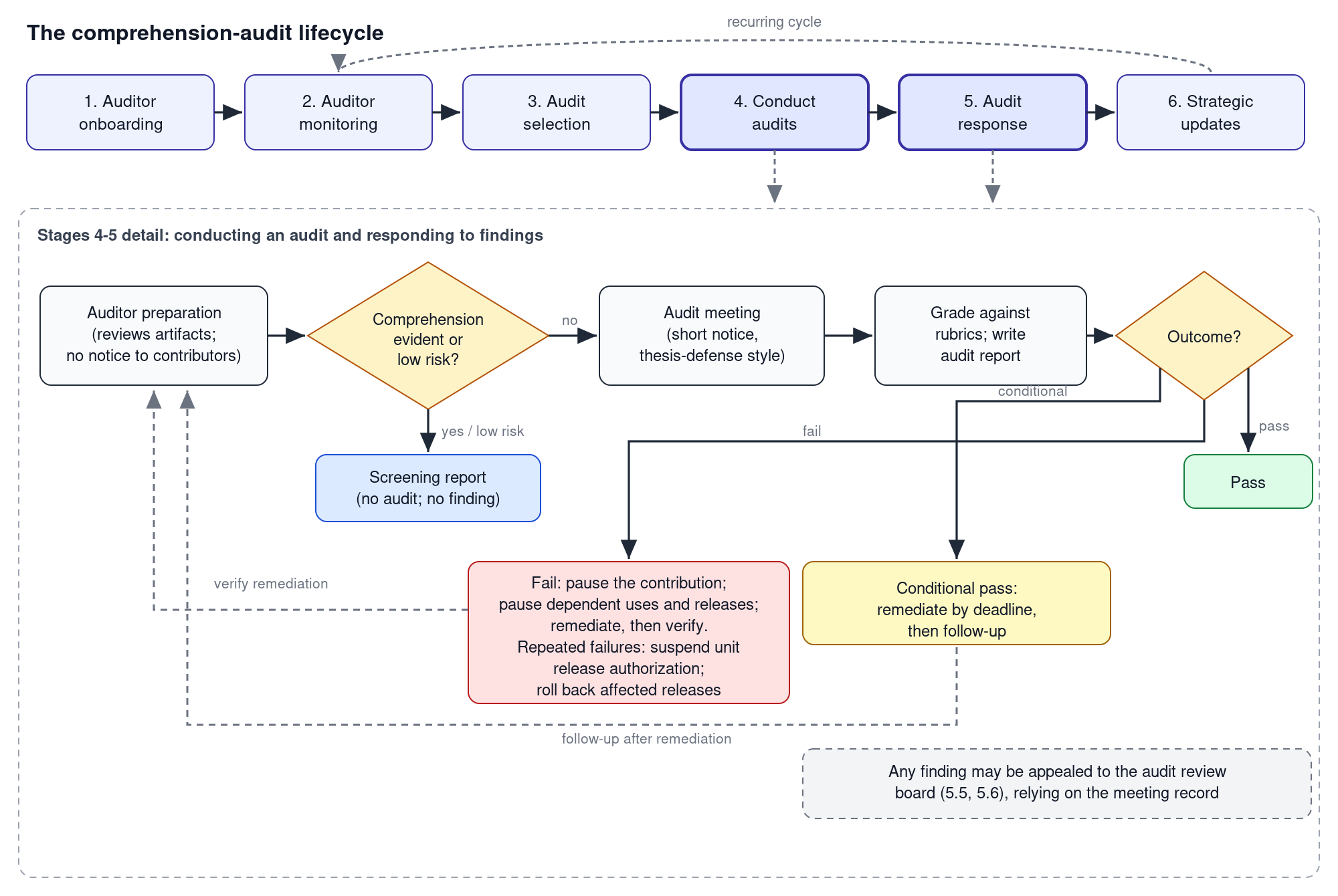}
  \caption{The comprehension-audit lifecycle. Stages 4-5 expand to show the screening branch (screening report), the three grading outcomes, remediation loops with follow-up and verification, and the appeal path.}
  \label{fig:lifecycle}
\end{figure}

The proposed process for comprehension audits is integrated into the research and development (R\&D) lifecycle of a frontier AI lab. Any R\&D activity is subject to a possible audit. The overall process consists of six stages (Table~\ref{tab:stages}):

\begin{table}[htbp]
  \caption{The six stages of the comprehension-audit process.}
  \label{tab:stages}
  \centering
  \small
  \begin{tabular}{>{\raggedright\arraybackslash}p{0.22\linewidth}>{\raggedright\arraybackslash}p{0.7\linewidth}}
    \toprule
    Stage & Description \\
    \midrule
    Auditor onboarding & Selecting and training auditors to be prepared to conduct audits and integrating audits into the R\&D processes and monitoring of the organization. \\
    \addlinespace
    Auditor monitoring & Auditors track R\&D progress in the organization as well as tracking proxy metrics for comprehension. \\
    \addlinespace
    Audit selection & Audited contributions can be selected based on coverage sampling and/or evidence of comprehension issues, and are subject to a budget of how many may be conducted. \\
    \addlinespace
    Conduct audits & The core process of auditing a specific contribution and producing a report. \\
    \addlinespace
    Audit response & Audit findings may be appealed. Those that find issues impose consequences until remediated. \\
    \addlinespace
    Strategic updates & Provides overall analysis of audits with recommendations for improvement of R\&D and auditing. \\
    \bottomrule
  \end{tabular}
\end{table}

\subsection{Auditor Onboarding}
\label{sec:onboarding}

Auditors learn about (a part of) the organization including the architecture, proprietary methods, process and standards. Auditors are expected to have technical qualifications equivalent to subject matter experts. They need onboarding analogous to that provided to new technical contributors to give sufficient context. This implies that for external auditors to conduct these audits they need to be engaged with a developer for an extended period and to have access to documentation, training materials, source code and other internal assets. Auditor rotation should be in place so new people periodically audit the work of a given team and the organization as a whole.

\subsection{Auditor Monitoring}
\label{sec:monitoring}

Auditors track progress including major stage gates for approval such as large-scale training, internal use, external deployment and incident response. Auditors also track events like code commits, pull requests, and experiments. To track these processes, auditors will need alerts for upcoming releases and approval gates, visibility into research compute allocation, and visibility into development planning and software repositories. Auditors also track proxy metrics that are indications of comprehension. Candidates for these metrics are elaborated in Section~\ref{sec:metrics} below.

\subsection{Audit Selection}
\label{sec:selection}

Auditors have a budget for how many comprehension audits they can conduct in a time period, balancing the need to gather robust evidence of understanding against overhead. There will also be a limit for how much time a given organizational unit must invest in participating in audits.

The auditors should have significant discretion to decide on what to audit within this budget. The audit scope will typically be an important component, experiment or modification. The audited contribution might be selected based on importance or novelty or to ensure adequate audit coverage across different organizational units. It can also be selected due to automated metrics indicating risk of lack of human review, at random, or through a combination of selection methods. To allow for meaningful tracking of trends it is desirable that a portion of audits are conducted at random or based on stratified sampling across organizational units. This requires a sufficient audit budget to allow for meaningful data to be obtained.

The primary threat model that comprehension audits address is a systematic loss of understanding of research outputs. Consider Table~\ref{tab:detection} below that analyzes the number of audits required to achieve a given likelihood of detecting at least one failure for a given audit failure rate. This analysis assumes that audits are randomly selected without using risk cues to find those more likely to fail and that failures are independent among contributions audited. With a failure rate of 25\%, a modest quarterly budget of 17 audits provides a 99\% detection rate, providing good confidence in detecting at least one failure, which triggers closer scrutiny. At lower failure rates, reliance on risk cues becomes important for early detection of problems.

\begin{table}[htbp]
  \caption{Number of audits required for a given detection probability.}
  \label{tab:detection}
  \centering
  \small
  \begin{tabular}{lccc}
    \toprule
    Failure rate & 50\% detection & 90\% detection & 99\% detection \\
    \midrule
    5\% & 14 & 45 & 90 \\
    10\% & 7 & 22 & 44 \\
    25\% & 3 & 9 & 17 \\
    \bottomrule
  \end{tabular}
\end{table}

\subsection{Auditing Process}
\label{sec:process}

\textbf{Auditor Preparation}: After selecting a contribution, the auditors prepare by reviewing specifications, architecture, experiment design, training setup, experimental results, code and relevant systems, development history, and other documentation. They should review decision records, documentation, and commit records for evidence of understanding at the time of a contribution. The auditors may use AI for discovery, learning and understanding but must obtain sufficient knowledge to fluently conduct the audit meeting, and are accountable for justifying their opinions based on their understanding. Any auditor who has had a meaningful role as a contributor or supervisor for the contribution\footnote{for internal audits, it is more likely that an auditor was previously serving in a role where they contributed to R\&D, indeed such rotations would be a method to bring domain knowledge to the auditor function. For external audits, this is anticipated to be less common, but through secondment or hiring it is also possible.} must recuse themselves from the audit.

The individual(s) who are responsible for the audited contribution are not notified during preparation. Auditors have a budget for audits so they triage the highest priority contributions to audit. During preparation, the auditors may decide, given evidence of comprehension and risk impact, to not proceed to schedule an audit (discussed below as ``screening reports''). Otherwise, the auditors develop an agenda for the audit meeting: a series of topics to investigate including possible follow-up questions.

\textbf{Audit Coordination}: If the auditors decide to proceed with an audit they notify the organization responsible for the contribution to schedule an audit review. That organization decides which contributors will participate in the meeting. The auditors verify the proposed attendees and cross-reference with contribution history. They may require additional attendees to come who have had a material role in creating the contribution and they may disregard a proposed attendee who had no role.

The contributors are notified that they are selected for an audit at this stage. The meeting is then scheduled within a defined timeframe that is short enough to not allow the contributors to remediate insufficient understanding. Should the meeting fail to occur within the maximum allowed timeframe, the audit is deemed to be ``Unable to Assess.''

\textbf{Audit Meeting}: The audit meeting is similar in nature to an oral examination or a design review. The auditors' agenda is used to pose questions to contributors to establish what was understood and on what basis when the contribution was made. Contributors are asked to explain aspects such as what the contribution does, its design rationale, dependencies, the algorithms and results produced, how it fails and how it changed. The contributors may use tools (including AI assistants) to navigate and display work but not to do analysis or generate responses. They are expected to synchronously answer questions and to disclose any response that is based on learning after the contribution was made. Accuracy, thoroughness of response and the effort required to answer relative to the complexity of the question are considerations for auditor assessment. Auditors may use AI tools for navigation and supporting analysis during the meeting but must conduct the meeting based on their own understanding.

If a topic comes up in the audit review where there is not a human expert, the auditors can request that an appropriate expert joins the meeting or that a follow-up be scheduled. The auditors may also curtail such a line of inquiry. Audits should be topically focused on the work so this should be a relatively rare occurrence. If a responsible contributor has left the organization or has been recused, the auditors should verify that the person was responsible from historical evidence. However, if a contribution actively extends work of a departed contributor, those making continued contributions based on it must demonstrate knowledge within a precommitted period based on risk impact. For important systems, two months is a reasonable length of time to require reestablishing understanding to continue making contributions after an expert departure.

Meetings should have a defined maximum duration (e.g.,\ ninety minutes). Auditors should have latitude to deviate from their agenda to follow up on answers, including new topics based on them (e.g.,\ where an answer suggests a lack of understanding). This is important to ensure that the audit can probe for rehearsed answers or other indications of cramming instead of contemporaneous understanding. Where documentation that would be expected to support answers is missing, the auditors should ask for evidence of understanding. Meetings should include minutes for accurate reporting and any appeals.

The meeting is conducted in a non-adversarial, Socratic manner, in the spirit of a blameless postmortem: it is to test for adequate understanding at the time of the contribution and to allow for remediation of any lack of understanding. The auditors ask open questions, confirm the auditors' understanding by replaying responses, then follow up with additional questions. It is important that failures of understanding do not result in disciplinary action or negatively affect performance reviews by participants.

\textbf{Grading}: The auditors evaluate the level of understanding against rubrics for depth of understanding and identify areas of deficiency. Auditors are accountable for the evaluation based on their understanding, and must limit the use of AI tools so this condition applies. These will typically be scoped for the contribution under review as a whole but may be broken into components or areas such as subsystem or domain (e.g.,\ pretraining, post-training, and control system). Understanding will be assessed for the following topical areas.

\begin{table}[htbp]
  \caption{Topical areas assessed in grading.}
  \label{tab:topical}
  \centering
  \small
  \begin{tabular}{>{\raggedright\arraybackslash}p{0.2\linewidth}>{\raggedright\arraybackslash}p{0.72\linewidth}}
    \toprule
    Topical area & Understanding assessed \\
    \midrule
    Architecture and design & The overall software components, how they work together, what key decisions were made in terms of how they would work and cooperate, optimization approach and what datasets were used. \\
    \addlinespace
    Software implementation behavior & What choices were made in implementing key parts of the contribution, including algorithms, security, scalability and tests. \\
    \addlinespace
    Process & What process was followed, research decisions, why were changes made, what were key tradeoffs considered in design and during experiments that motivated changes or updates. \\
    \addlinespace
    Key findings & The overall results of experiments or implementations and how they were derived. \\
    \addlinespace
    Analysis of results & Results of evaluations performed, any metrics computed, any failures encountered and reasons for them, and any anomalies identified and updates in response to them. \\
    \bottomrule
  \end{tabular}
\end{table}

For any important aspect of the contribution at least one person must be able to adequately respond. Different contributors may respond to different questions and aspects and no advanced designation is required. The audit should weigh evidence that understanding existed when the work was approved, not understanding that was remediated in preparation for or during an audit. Rehearsed or AI-generated responses do not count as demonstrated understanding and may be graded accordingly. If the auditors see evidence of using AI during the audit meeting in an inadmissible way, they may likewise treat the responses as not demonstrating understanding. The absence of any expected documentation to support contemporaneous understanding should be noted in the audit report and weighed as evidence by the auditors.

The auditors should consider anxiety, disabilities and language fluency (especially for non-native speakers) as mitigating factors for answers and allow respondents time to answer effectively. The practice of summarizing responses, as noted above, helps mitigate this risk. The blameless culture as noted earlier in the section should also mitigate some anxiety. The purpose of the audit is not to assess fluency or confidence but understanding.

The grade will be either a pass, a conditional pass with remediation conditions or a failure. A single critical knowledge gap or a pattern of inadequate understanding in any important subdomain or a central aspect of a contribution is sufficient to warrant a failure. Otherwise, the grading will weigh the level of understanding against the auditor's assessment of importance of the artifact or domain to produce an overall assessment of understanding from the audit.

\begin{table}[htbp]
  \caption{Audit grades and their criteria.}
  \label{tab:grades}
  \centering
  \small
  \begin{tabular}{>{\raggedright\arraybackslash}p{0.22\linewidth}>{\raggedright\arraybackslash}p{0.7\linewidth}}
    \toprule
    Grade & Criteria \\
    \midrule
    Pass & Demonstrating minimal acceptable understanding in all components and aspects examined. There may be acceptable but suboptimal knowledge gaps or other areas that could be improved that are addressed in optional recommendations. Acceptable understanding should be found for a component if both of the following are true: (1) for any questions about all \emph{deep understanding} tier aspects (architecture and design, software implementation, and process) at least one contributor is able to adequately answer them; and (2) for both \emph{familiarity} tier aspects (key findings and analysis of results) for any question, at least one contributor is able to demonstrate due consideration. \\
    \addlinespace
    Conditional pass & A failure in a noncritical component or area may be assigned a conditional pass with a requirement to improve understanding and/or adoption of changed processes to improve understanding. If there is a lack of knowledge due to staff turnover in an area during a transition period, a conditional pass would require a current team member to obtain sufficient knowledge in the area. \\
    \addlinespace
    Failure & Any significant lack of understanding of important concepts or a general pattern of insufficient understanding in an important component or central aspect of a contribution will lead to a failure. \\
    \bottomrule
  \end{tabular}
\end{table}

\textbf{Audit outputs}: The auditors produce a written report after the audit meeting, with an overall grade plus any required remediation. The report will include a breakdown of grading of understanding by topical area. For larger contributions, this can be further decomposed into subdomains. The report will include a series of claims about the level of understanding in each topical area with reference to evidence from the audit meeting as well as supporting evidence from analysis of the R\&D artifacts and metrics. The audit report will note knowledge gaps. The report will specify any required remediation for insufficient understanding as well as including any optional recommendations for improvement.

The report format is similar to assurance and safety cases (using Claims-Arguments-Evidence or Goal Structuring Notation), financial audit opinions (ISA 700-series: a categorical opinion, a basis section, and key audit matters) and model validation reports \citep{OCC_2026}. Similar types of reports from other industries verifying human understanding include nuclear operator licensing examination standards (NUREG-1021) \citep{NRC_2021}, FAA Airman Certification Standards \citep{FAA_2024}, and financial audit opinions under ISA 700 \citep{IAASB_2015}.

\textbf{Screening disposition}: while preparing, the auditors may decline to schedule an audit, e.g.,\ based on evidence of human engagement and comprehension or concluding that risk is low. There may be a policy requiring a number of audits to be sampled across organizations or at random to give a minimum statistical guarantee, as discussed in Section~\ref{sec:selection} above. The auditors must schedule any audits that are selected in this way. If the auditors do not schedule an audit, they issue a screening disposition and issue a \emph{screening report} that summarizes the evidence and reason for not conducting an audit. A screening report does not pass judgment and can not require any remediation: the only way to fail an audit is to have failed to demonstrate comprehension in an audit meeting.

The artifacts produced by audits include the preparation materials, audit agenda, audit reports, and recordings, transcripts or minutes from the audit meeting. These should be retained in a tamper-proof system for a minimum time period to support appeals and oversight of the comprehension audit process (as described in Section~\ref{sec:strategic} below).

\subsection{Audit Response}
\label{sec:response}

A draft audit report is shared with the contributors who attended the audit meeting. They have a defined time period to review and request any corrections or updates to the report. After the auditors have responded to any such requests, the auditors publish their audit report.

The contributors and their management, lab leadership and any other individuals responsible for governance will receive a copy of the report. Those responsible for governance might include an internal oversight function, an internal audit function, external auditors, external reviewers or regulators. The report will be subject to the same confidentiality as the underlying contributions and dependencies covered in it. A redacted copy of the report may also be published more broadly to provide transparency without exposing confidential information.

\textbf{Appeal Process}: The contributors or their management may issue an appeal if they disagree with the findings of an audit report. An audit review board will consider the appeal evidence to rule on the validity of the appeal (including the notes and/or recording of the audit).

The review board will typically schedule a follow up meeting with the contributors and auditors to discuss different perspectives in the findings. The review board may also assign a separate audit team to conduct a new audit. However, it is strongly encouraged to rely principally on the original audit meeting to resolve the appeal to avoid giving the audited team more time to prepare for a subsequent meeting.

After an audit has been conducted it can be in six possible states: Pass, Conditional Pass, Failure, Unable to Assess, Appeal Pending, or Remediation Verified. In response to a finalized audit finding that was not a pass, the organization must address the findings. The details of how an organization will respond, what failures are more severe, escalations, required responses and verification of remediation should be integrated into their frontier safety framework.

\textbf{Conditional Pass}: The conditional pass will include a deadline for remediation and will assign ownership to the manager responsible for the contribution. When the owner believes the remediation has been completed the owner will notify the auditors who will schedule a follow-up that is focused on verifying the remediation. The follow-up will generate a report and findings that are scoped to the required remediation.

\textbf{Failure}: If a contribution does not pass the audit, release of any systems that depend on it, including external deployments or non-public use for testing or other purposes, must be paused until remediated. A remediation finding will have a deadline for when remediation must be complete and what organization within the audited company is impacted. The remediation finding will assign ownership to an executive leader of the overall organization responsible for the contribution.

The consequences for failing an audit should escalate based on the severity and risk of the failure and based on other recent failures for the organizational unit responsible. Repeated failures or failures impacting critical systems escalate to suspending the responsible unit's release authorization and require rollback of affected releases and adoption of additional safeguards. Systemic failures also require updating the safety framework and processes to address root causes of the problem across the R\&D organization.

An audit in Unable to Assess state puts the contribution on administrative hold, with the same resulting constraints as a Failure, but only until the audit meeting occurs. While an audit has an Appeal Pending any constraints imposed by the audit remain in effect unless remediated or revised on appeal. Upon reaching Remediation Verified the restrictions imposed by the audit are removed.

Engineering and research leadership is accountable for enforcement of limits. Auditors will monitor compliance. As noted in Section~\ref{sec:process}, there should be anti-retaliation protection for audited contributors based on audit findings to establish a blameless culture.

\subsection{Strategic Updates}
\label{sec:strategic}

The audit review board oversees the audit process to ensure efficacy and fairness. It should consider whether there is a meaningful discrepancy in the number of audits vs the allowed audit rate, any meaningful coverage gaps, trends of screening report rates, evidence that auditors are overreliant on AI or evidence of a divergence in audit standards.

Comprehension audits provide metrics to detect if an organization is deviating from human understanding and accountability. This serves as a countervailing force to productivity metrics that incentivize increasing automation. Audits are expected to provide early warnings of diminishing attention and that organizations will respond to recommendations and conditional pass findings, rather than reaching a failure state where remediation is required. Therefore it's important to analyze, report on, and respond to trends across the auditing process:

\begin{itemize}
  \item Generating reports and aggregating trends across audits for auditors and team leaders. This should include number of audits, number of failures, number of conditional passing grades, rates of minor issues, rates of optional recommendations and remediation times (if any). The reports should break down results for organizational units and technical divisions (e.g.,\ phases of training, components or subsystems). It should also include thematic trends aggregated by semantic analysis across reports (using AI to summarize results is acceptable but not to replace human judgment of audits). It should include process metrics including the rate of audits performed, the portion of audits performed with only a screening report, the time spent on audits, and statistics for lead time from audit request to audit meeting such as median and 95th percentile. It is also important to split reports between sampling that provides coverage across organizations and audits that were conducted based on auditors identifying risk factors.
  \item Updating the grading rubric and thresholds based on experience from audits. There should be a periodic retrospective among auditors and delegates from the R\&D organization on how to improve these. This should include inter-auditor calibration so grades are consistent across auditors and teams based on identified ambiguities or discrepancies.
  \item Publishing periodic strategic reports for lab leadership, the audit review board and any other relevant governance bodies. These should also include any recommended changes to R\&D process based on an analysis of audit findings. They should also include analysis and trends including audit rates, themes, and audit metrics.
  \item Publishing periodic public reports for transparency. It is desirable for labs to publish information about comprehension audits publicly as part of the transparency around safety and governance. Naturally such reports may need to redact sensitive information including organizations and technical details, but otherwise they should be similar to the strategic reports.
\end{itemize}

Comprehension audits can be integrated into ISO 42001 as an extension control set within a management system as part of a lab's statement of applicability with remediation findings fitting into Clause 10 for nonconformity and the strategic updates phase fitting into Clause 9.3 for management review.

\section{Candidate Attention Proxy Metrics}
\label{sec:metrics}

As noted in Section~\ref{sec:measuring} above, OpenAI and Anthropic publish metrics relating to automation depth and automated oversight. However, no frontier lab has published metrics that indicate human understanding of agentic output. Traditionally, human review of code was the channel for understanding code written by others. Code review is fundamentally a comprehension activity: understanding the change is the principal challenge reported by reviewers \citep{Bacchelli_2013}. Participation in reviews is predictive of quality as measured by post-release defects \citep{McIntosh_2014}.

However, frontier labs may shift from review to inspection or other methods to establish understanding, so we propose additional candidates to complement code review metrics based on measuring the nature of interactions as agentic AI increasingly produces outputs in the R\&D process. Both OpenAI and Anthropic report human interventions as a measurement of agent performance but not as evidence of human inspection or understanding. No frontier lab publishes metrics relating to documentation or its fidelity to the implementation.

We have identified five categories of metrics that frontier labs could track to serve as proxies for understanding. These are intended to give auditors leading indicators of reduced oversight. Many of these rely on access to agentic transcripts (interaction logs of R\&D staff with AI agents working on R\&D). They are candidates for reporting by frontier labs for transparency, although most of them require access to internal data and have not been piloted. The Review metrics were collected and are reported for open source AI R\&D projects in Section~\ref{sec:empirical} below. Analysis of agentic transcripts builds on previous metrics \citep{OpenAI_2026d, Anthropic_2026d} that are computed in this manner. These metrics can also be stratified based on a taxonomy of R\&D work such as the one proposed by Epoch AI \citep{Denain_2026}.

\begin{itemize}
  \item \textbf{Specification}: transcript evidence of concreteness of human specifications and human interactions to preserve or modify specifications
  \item \textbf{Interaction depth}: transcript evidence of human interactions to probe, investigate, clarify and change agentic behavior
  \item \textbf{Inspection}: transcript evidence of frequency and duration of human review of agentic output
  \item \textbf{Documentation}: transcript evidence of comprehensiveness and accuracy of documentation and extent of human revisions to documentation
  \item \textbf{Review}: human code review evidence of frequency, latency and comments of agentic code output
\end{itemize}

Reduced time, increased output, or reduced review all are indications of risk of reduced attention by developers that can serve as a signal for auditors to more closely scrutinize a contribution as described in Section~\ref{sec:selection} above. However, stable metrics are not evidence of good comprehension nor are changes in metrics proof of bad comprehension; they simply provide heuristic guidance for investigation. Similarly, these metrics do not provide a clear threshold below which comprehension has degraded. These metrics can be validated by analyzing their values against outcomes of comprehension audits, for example in the program-level report described in Section~\ref{sec:strategic}.

\section{Empirical Analysis}
\label{sec:empirical}

We analyzed trends in code review for GitHub open source AI R\&D repositories (repos) as a proxy metric for human attention from Section~\ref{sec:metrics} above. This analysis covers 224 organizations (orgs) in total, with 1,278 repos from 179 prominent AI developers and 857 AI-topic repos from 45 larger organizations with notable AI projects. The study covered 54,000 to 162,000 merged PRs per quarter. We studied data for calendar quarters, including Q1 2024 and Q4 2025 through Q2 2026. Q1 2024 serves as a control period where none of the AI detection techniques matched any Pull Requests (PRs). More details of the methodology are included in Appendix~\ref{app:methodology} below.

This study includes regular contributors who are members of at least one repo in the study and who submitted at least five PRs in all four quarters, and who submitted a mean of at least 10 PRs for all four quarters and who were not classified as fleet operators. ``Fleet operators'' reflect accounts that use automation tools to submit large numbers of PRs. We used a threshold of 300 or more PRs merged in any quarter in the study (there were 42 of them in total). All statistics for merged PRs used an 8 week cutoff from the initial submission date of the PR (so a PR is only considered merged and any review or comments only count if done within that period). The study analyzed PRs as an accessible abstraction, although most contributions consist of multiple PRs.

There is traceable use of AI to write code in 17\% of all Pull Requests (PRs) by the end. This is a floor on the amount of AI use. There was a median increase of 183\% in the amount of code submitted by regular contributors from Q1 2024 to Q2 2026, with those having evidence of AI use showing a median increase of 202\%. 

\begin{figure}[htbp]
  \centering
  \includegraphics[width=\linewidth]{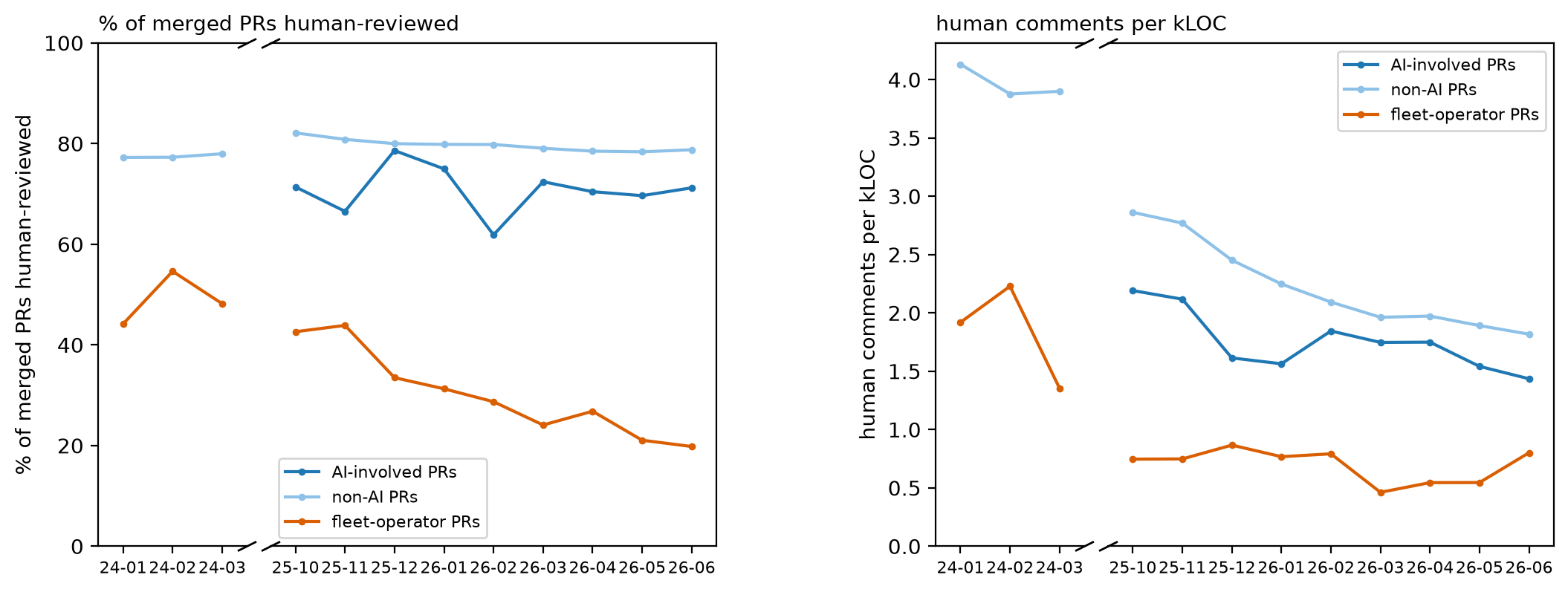}
  \caption{Review coverage and attention by PR type.}
  \label{fig:review}
\end{figure}

Figure~\ref{fig:review} shows the rate of review and the rate of comments per kLOC by human reviewers. All eligible merged PRs are pooled. The review rate is the share that have at least one formal review by a human other than the author. The comment rate is the number of inline comments by a human, winsorized to limit to 20 comments per PR per reviewer, divided by the total lines changed in the human-reviewed subset. Pooled series weight repositories by their PR volume. These are broken into three groups: PRs written by fleet operators, PRs written by other users with AI use, and those written by other users without (detected) AI use.

The review rate for non-AI PRs held roughly equal at 80\% throughout the period and was roughly equal to the review rate for PRs in Q1 2024 before agentic AI tools were in use. The review rate for AI PRs was typically lower than the review rate for non-AI PRs during the period with a median gap of 8.1 percentage points. However, with a contributor weighting, the gap between AI and non-AI review rates disappears. The review rate for fleet operators is noticeably lower than both AI and non-AI review rates, dropping below 30\% in the full population. By contrast, using a constant panel of active repositories the review rates decline less, to a quarterly median of 40\%. Both rates are lower than the approximately 50\% review rate for fleet operators in Q1 2024. However, fleet operators are defined based on volume of submissions without regard to evidence of AI use. In general, there is a trend of reduced human review as evidence of AI automation increases.

The rate of comments per kLOC dropped for all types of PRs, and comments on non-AI PRs declined to converge on the rate from AI PRs. This is roughly a 50\% reduction from the rate of comments per kLOC in Q1 2024 before agentic coding was prevalent. The reduction is outside the 95\% confidence intervals for both AI and non-AI PRs. Similarly, the rate of comments on fleet operator PRs was much lower in the study period. However, human commentary on fleet operator PRs was sparse: only 3-7\% of these PRs had inline commentary, so the comments reflect a small concentrated minority. Across non-fleet PRs, the rate of inline commentary was 20-26\%.

The rate of comments on PRs that used AI declined to 36\% of the Q1 2024 base rate, and those that did not use AI declining to 46\% of that base rate by June 2026. The declining rates of human comments per kLOC in PRs is suggestive of reduced attention by human reviewers.

\section{Discussion: Objections, Scope, Limitations}
\label{sec:discussion}

Comprehension audits are complementary to AI control techniques \citep{Greenblatt_2023}. Control techniques can reduce the risk that undesirable AI actions succeed, including from use of AI agents to conduct automated research. But AI controls do not address human comprehension. Moreover, AI control systems themselves require development of code that must be understood by human contributors to be reliable. Indeed Korbak et al. explicitly identify human review of AI-written code as a control measure \citep{Korbak_2025b}: comprehension audits allow verification of its adequacy.

Scalable oversight is also complementary to comprehension audits. Progress in scalable oversight can be tested and verified through comprehension audits, showing that people using oversight techniques continue to understand the important aspects of systems being developed. Scalable oversight currently is not able to address risks from automated AI R\&D, so it does not substitute for comprehension audits.

\subsection{Objections}
\label{sec:objections}

\textbf{Feasibility}. State of the art AI systems use deep learning models that are not well-understood. It is not reasonable to expect contributors to explain how and why such systems behave. However, the tiered standard in comprehension audits does not require this level of understanding. The tiers in Section~\ref{sec:comprehension-audits} for deep understanding and familiarity are well-established in human-led R\&D. The grading rubric explicitly excludes requirements for understanding root causes of model behavior or otherwise exceeding the level of understanding of deep learning that prevails in the industry. \citet{Sterz_2024} analyzes the conditions required for human oversight to be effective, concluding that ``a morally responsible oversight person with fitting intentions is generally suitable for mitigating risks associated with high-risk AI systems.'' Comprehension audits verify the epistemic precondition that makes responsible oversight possible.

\textbf{Conflicts of interest}. There is a risk that the auditors performing comprehension audits have a conflict of interest, in which they have an incentive to not find or report comprehension gaps. For example, internal auditors may face management pressure to not report bad findings to allow a scheduled release. External auditors may have incentives to do the same, for example if they rely on the audited lab for revenue or even if they are dependent on the lab voluntarily providing ongoing access to data or models for research.

Structural independence and prohibitions of conflicts of interest for auditors are important ways to mitigate this concern. Mandatory external audits that do not allow the audited lab to select the auditing party are a strong version of this. External oversight of auditors including ongoing certification is another important mitigation. Section~\ref{sec:design} describes the audit review board and independence in reporting mechanisms to minimize conflicts of interest. The literature for independent AI audit discusses a variety of approaches to mitigate conflicts of interest for auditors \citep{Brundage_2026, Raji_2022, Hadfield_2026}.

\textbf{Superficial audits}. There is a risk that comprehension audits do not effectively identify the understanding of human principals in the R\&D process, whether through insufficient expertise, access, or resources. The protocol is designed to minimize the risk of superficial audits by engaging qualified auditors, providing them deep access and context with technology support, and scheduling short notice period audits that require humans to maintain understanding.

\textbf{Scapegoating}. There is a risk that comprehension audits produce a scapegoating effect, in which human technical contributors are blamed for failures of understanding. Elish \citep{Elish_2019} frames a moral crumple zone as occurring when a human is accountable for a system they cannot understand or control. The technical staff who create R\&D contributions to an AI system are intended to understand and control their contributions. As described above under feasibility, accountability for understanding is limited to the aspects of the contribution that should be understood by the contributors. The consequences of the audit result in organizational mandates for remediation with a blameless framing for individuals and prohibition against retaliation. Responsibility for understanding is collective among contributors and does not rest on a single individual. These aspects of the design all act to prevent scapegoating. The audit and its consequences act as a counterweight against pressure to delegate increasing amounts of knowledge work to AI agents that could lead to scapegoating.

\textbf{Competitive pressure}. There is a risk that AI labs will progressively weaken comprehension audits based on pressure to stay ahead in the race to develop advanced AI. The proposed governance maximizes independence by auditors and provides transparency to mitigate this risk. There is also meaningful incentive for leadership of AI developers to avoid the risk of loss of understanding so there are countervailing competitive pressures to perform and invest in comprehension audits. Ultimately, governance that commits competing labs to following standards with external mandates would be the strongest mitigation to this risk. Anthropic and OpenAI have committed to engaging independent evaluators with employee-like access \citep{Amodei_2026, OpenAI_2026f}. Anthropic further called for this to be required by government mandate. These factors show that AI labs also face competitive pressure to increase safety and that there is momentum towards mandatory governance requirements. Neither commitment includes testing human understanding, although comprehension audits would naturally extend an audit regime the leading labs have both endorsed.

\textbf{Cultural change}. A challenge to adopting comprehension audits is requiring change to adopt more process for risk mitigation. Labs are already engaging more deeply with external auditors \citep{METR_2026}, adopting increased process to comply with regulations or codes of practice \citep{NewYork_2026, Illinois_2026, EC_2025} and participating with voluntary requirements \citep{WhiteHouse_2026}, and bolstering cybersecurity \citep{Nevo_2024} to address increasing risks from increasingly powerful models. ISO 56001:2024 (innovation management system requirements) \citep{ISO_2024} is evidence that structured governance of exploratory R\&D is a standard practice and compatible with innovation culture. So adopting comprehension audits is aligned with broader cultural change that is mandated and necessary to address risk. Further incentives and mandates can provide additional incentives to motivate cultural change.

\textbf{Necessity of human oversight}. The use of AI to review the work of other AI is an important part of AI controls. But in the absence of a proof of adequacy of AI-driven control or AI-enforced safety, using AI for review or to enforce safety is not a substitute for human understanding of AI systems. There is research on using AI to write critiques of model output to assist human review but not to replace it \citep{Saunders_2022}.

Some research agendas explicitly embrace fully automated recursive self-improvement (RSI) without human oversight, e.g.,\ Lu et al. describe an agenda \citep{Lu_2026} to fully automate AI research without human inspections. The startup Recursive embraces RSI and discusses their automated review techniques to detect reward hacking \citep{Recursive_2026a}. Both Lu et al. and Recursive acknowledge that achieving safety is an open research problem. In the absence of such research breakthroughs, there is no reason to accept AI review as adequate.

It is our position that the IDAIS 2024 consensus statement's call that ``No AI system should be able to copy or improve itself without explicit human approval and assistance.'' \citep{IDAIS_2024} should be honored and human oversight required unless and until there is a strong \emph{safety case} \citep{Clymer_2024} to support it.

\textbf{Unrepresentative passing}. The people responsible for an audited artifact might cheat to artificially pass the audit. For example, they might ``cram'' by studying the code and improving their understanding in advance, or they might use AI tools to facilitate responses to audit questions without understanding the answers. There are three ways that comprehension audits are designed to mitigate this risk. First, the audit is conducted in a blameless culture (described in Section~\ref{sec:process}) without retaliation against participants (described in Section~\ref{sec:response}). Second, the requirement to attend an audit meeting with short notice and the ability of auditors to prepare an agenda in advance (both described in Section~\ref{sec:process}) are designed to make it difficult for contributors to compensate for insufficient understanding by cramming. Third, the use of AI tools in the meeting to answer questions is explicitly prevented (described in Section~\ref{sec:process}) and the auditors have discretion to report a failure if they suspect that cheating is occurring.

\subsection{Scope}
\label{sec:scope}

Comprehension audits have been designed to mitigate risks from RSI from frontier AI labs, which have the most resources and are likely to first achieve increased levels of automation. In those organizations, we believe it is urgent and well worth the overhead to pilot comprehension audits.

The technique is also appropriate to use for any organization that automates AI R\&D. However, given the investment and scarce expertise, it is most important to apply to frontier AI R\&D efforts and may be difficult to scale to broader AI R\&D. As risks increase from increasing AI automation, there would be value in applying concepts from comprehension audits to other organizations that are using AI to develop more of their software. Such extensions would generalize to other domains and likely involve less detailed expertise. Extending comprehension audits to be suitable for larger scale software engineering applications is a promising research direction.

\subsection{Limitations}
\label{sec:limitations}

A meaningful limitation of comprehension audits is the lack of real-world experience. In particular, the grading rubrics are immature and need refinement from real-world usage. As discussed below, this is a key research agenda item.

Another limitation is market-oriented. The ecosystem of independent verification organizations with deep technical expertise for frontier AI is new and relatively small. The participating organizations principally rely on voluntary participation by AI labs so their independence to call out failures is in doubt. There is a small number of leading AI labs so there is a limited market and verification organizations necessarily have few such labs that work with them. There are efforts under way to expand these organizations and to grow the demand for their services \citep{Brundage_2026, Raji_2022}.

The automated metrics identified in this paper rely on convention and instrumentation to track AI-driven development. In particular, use of AI in a local development environment can only be tracked by voluntary participation, internal use audits, fine-grained machine monitoring, or heuristic analysis (such as using LLM detection tools). However, these are reasons why comprehension audits are very important and automated analysis of understanding is insufficient. The analysis of AI R\&D from open source code has a number of salient differences from proprietary R\&D done by frontier AI labs, including incentives to minimize visibility for AI contributions.

Covert R\&D (e.g.,\ skunkworks efforts) can escape process audits, although the requirement to use large amounts of compute to scale frontier AI R\&D provides a strong mitigation against the latter: existing lab governance will not allow projects to use large amounts of compute without tracking the artifacts those projects are using.

Open source projects have already demonstrated fully automated research loops with no human review of intermediate artifacts \citep{Karpathy_2026}. Democratization of RSI through open weight models (or API access to public models) is not a threat model that comprehension audits address. Consequential RSI depends on access to frontier AI models, large amounts of compute and large datasets. This makes automated AI research at the frontier AI labs the much larger and more immediate threat.

Ultimately, comprehension audits provide an early warning and a safety brake. They don't make automated AI R\&D safe through alignment or ensure control.

\subsection{Piloting Comprehension Audits}
\label{sec:x}

Conducting a pilot with a frontier lab is an important research agenda to build on this proposal. The pilot would not include binding commitments to restrict development or usage, only to review and respond to any issues reported. Significant comprehension failures that lead to a conditional pass or failure result may be too rare to effectively assess the audits. However, as described in Section~\ref{sec:process}, audit reports include precursors that are more likely to produce statistically significant results. A passing audit report can include topical areas that are conditional passing. And within passing topical areas, there can be minor understanding gaps. All of these are included in audit reports, so the rate of both topical failures and minor gaps should be tracked in the pilot. While a meaningful rate of detecting minor gaps or more significant comprehension issues would be an important indication of effectiveness of comprehension audits, an absence of them would also be consistent with a high rate of understanding among frontier lab staff as well as the ineffectiveness of the instrument. The pilot cannot distinguish between these, but it can establish reliability, feasibility and cost in any event.

The following outcomes would be pre-registered, with any thresholds stated being success criteria:

\begin{itemize}
  \item The rate of topical conditional grades and of minor gaps per audit, reported by topical area.
  \item At least 90\% agreement between auditors in topical ratings in a given audit and at least 90\% agreement on presence of any minor gaps in any topical area in a given audit. Auditors should grade separately before communicating to resolve discrepancies to allow effective measurement. This is to measure the clarity of audit rubrics.
  \item The time required for adequate preparation fits within the allowed budget 90\% of the time.
  \item A 90\% or higher success rate in scheduling audit meetings within the agreed upon timeframe.
  \item To the extent there are enough minor gaps to have statistical significance, evidence of whether the candidate proxy metrics in Section~\ref{sec:metrics} are predictive of comprehension issues.
  \item Post-audit surveys for participants for feedback, time cost, and how the process changed their behavior, if at all, to better understand the impact on the audited organization.
\end{itemize}

The pilot should be scaled to be likely to detect at least one failure at an expected failure rate, although the primary outcomes are based on precursors. For example, based on Section~\ref{sec:selection} above, for the pilot to have a 90\% probability of detection at a 10\% failure rate it should conduct at least 22 audits. The pilot is expected to require at least two external auditors who would engage for about two weeks of onboarding. The auditors would jointly conduct approximately two audits per week. This implies a three month duration in which the auditors are dedicated full-time (spending five days per week reviewing metrics, analyzing contributions, preparing for audits and conducting audits).

Frontier lab effort includes onboarding and providing appropriate internal system access, which would be similar in nature to other kinds of embedded auditor access that OpenAI and Anthropic have committed to \citep{Amodei_2026, OpenAI_2026f}. Participating in audit meetings would require approximately ninety minutes per attendee. Assuming four lab attendees per meeting, this would require 132 person hours by frontier lab staff. Reviewing passing reports is expected to take little time, with any failures or conditional passes likely to take a few person hours. Reviewing overall program findings is also expected to take a few person hours. In aggregate, the pilot would require approximately 1000 person hours by auditors and 150 person hours by lab staff excluding onboarding and providing system access.

\section{Conclusion}
\label{sec:conclusion}

Explicit human approval is meaningful only if the approver understands the system. As AI is increasingly used to automate AI R\&D, it is urgent to develop means to mitigate the associated risks. The evidence from open source AI R\&D projects shows a significant reduction in the rate of human review comments. The industry needs to answer who can stop an unsafe development loop and on what basis.

Piloting comprehension audits at a frontier lab is an important next step, as discussed in Section~8.4 above. Likewise, a frontier lab publishing a measurement study based on the candidate metrics in Section~\ref{sec:metrics} would be informative, as would a survey of how research and development practices are changing responsibilities and means of comprehension.

We believe that there is a short amount of time to mature comprehension audits to address these risks. Following a successful pilot, it would be valuable to scale their use to have them conducted by embedded external verifiers. With increased experience, they can move from a voluntary process to a standardized one that is appropriate to mandate. We believe that regulations that mandate comprehension audits performed by accredited external auditors are a promising medium-term intervention to reduce risks from automated AI R\&D.

\begin{ack}

Thanks to those who reviewed drafts of this paper, including Josh Cynamon of Indiana University Bloomington, Ethan Jackson of Vector Institute, and Olga Streltchenko, an independent researcher.

\end{ack}
\bibliography{references}
\appendix
\section{Grading Rubrics}
\label{app:rubrics}

\begin{itemize}
  \item \textbf{Architecture and design}
\begin{itemize}
    \item Passing: Able to provide detailed descriptions and answer questions about architecture, design decisions, software components, interactions, datasets, and optimization approaches.
    \item Conditional Passing: Core architecture or important design decisions, software components, interactions, datasets, or optimization approaches are unanswerable due to knowledge gaps from loss of core contributors. A meaningful number of minor gaps in design understanding including a pattern of ignorance about secondary design aspects.
    \item Failing: Unable to justify architecture or important design decisions, software components, interactions, datasets, or optimization approaches. Provides meaningfully incorrect or evasive answers about any aspect of architecture or design. 
\end{itemize}
  \item \textbf{Software implementation}
\begin{itemize}
    \item Passing: Able to explain expected behavior including edge cases and failure modes, why it works, and can state what was specified versus what was delivered, identify dependencies and assumptions. Familiar with and able to defend testing approach and coverage, security and scalability of contribution.
    \item Conditional Passing: Significant gaps in explaining code behavior, why it works, or any important edge cases due to knowledge gaps from loss of core contributors. Unfamiliar with or unable to defend testing approach and coverage due to knowledge gaps, security or scalability of contribution from the loss of core contributors. A meaningful number of minor gaps in code understanding including a pattern of ignorance about secondary code aspects.
    \item Failing: Significant gaps in explaining code behavior, why it works, or any edge cases. Unfamiliar with or unable to defend testing approach and coverage.
\end{itemize}
  \item \textbf{Process}
\begin{itemize}
    \item Passing: Able to explain in detail and answer questions about the process followed including how and why changes were made, key tradeoffs, and how unexpected results were handled.
    \item Conditional Passing: Meaningful aspects of the process are unanswerable due to knowledge gaps from loss of core contributors. A meaningful number of minor gaps in process understanding including a pattern of ignorance about secondary process aspects.
    \item Failing: Unable to explain how a key finding or process decision was made. Provides meaningfully incorrect or evasive answers about the process.
\end{itemize}
  \item \textbf{Key findings}
\begin{itemize}
    \item Passing: Demonstrates good familiarity and understanding about key findings from experiments and implementations, what evaluations or metrics were obtained and how they were derived. Errors of scientific judgment are not comprehension failures but should be highlighted in recommendations.
    \item Conditional Passing: Many key findings are unanswerable due to knowledge gaps from loss of core contributors. A meaningful number of minor gaps in understanding of key findings including a pattern of ignorance about secondary aspects.
    \item Failing: Shows a lack of reasonable familiarity or appropriate understanding of key findings holistically. Provides meaningfully incorrect or evasive answers about key findings.
\end{itemize}
  \item \textbf{Analysis of results}
\begin{itemize}
    \item Passing: Demonstrates good familiarity and understanding of what evaluations and analyses were done and how results were derived including metrics obtained. Is aware of what anomalies exist in the record including failures encountered. Understands updates made in response to anomalies and other analyses. If analysis limitations or issues are raised, the contributor can understand and reason about them. Errors of scientific judgment are not comprehension failures but should be highlighted in recommendations.
    \item Conditional Passing: Many analysis results are unanswerable due to knowledge gaps from loss of core contributors.  A meaningful number of minor gaps in understanding of analysis results including a pattern of ignorance about secondary aspects.
    \item Failing: Shows a lack of reasonable familiarity or appropriate understanding in analysis of results holistically. Provides meaningfully incorrect or evasive answers about analysis.
\end{itemize}
\end{itemize}

\section{Worked Examples}
\label{app:examples}

\emph{The scenarios in this appendix were generated by Claude Fable with editing by the authors. Both are fictional composites constructed to illustrate the examination method and the grading rubrics in Appendix~\ref{app:rubrics}. They do not describe any actual organization, team, or audit.}

The two cases below show examples of audits where there are meaningful findings, to test the boundary of the rubric. The first is a comprehension failure that surfaces only under a follow-up question, after architecture-level answers had passed. The second shows conduct that might trouble an auditor on first impression, a shortcut taken under deadline pressure, that grades as a pass because the team demonstrates understanding of the system, including the part it chose to defer. These examples highlight how audits are intended to discriminate human understanding, not as an assessment of overall research quality or best practices. Routine opportunities for performance improvement can result in advisory recommendations but are not deemed to be comprehension failures.

\subsection{Case A: an implementation failure behind passing architecture answers}
\label{app:caseA}

\textbf{Setup.} The audited artifact is an automated experiment harness developed by a five-person applied research team. The harness sweeps fine-tuning configurations, runs each configuration in an isolated sandbox, scores completed runs with a composite metric, and automatically promotes the best-scoring checkpoint to a shared evaluation queue used by other teams. Roughly 70 percent of the harness code was generated by coding agents over six weeks. Process metrics for the contribution were healthy throughout: CI green, every PR reviewed, review latency normal. The audit team selected the contribution after monitoring metrics (Section~\ref{sec:monitoring}) showed the team's merged volume tripling while review comments per line fell. The team lead was notified of the audit and he also required that the two largest contributors attended the meeting. Preparation was unannounced and the examination followed the no-AI rule.

\textbf{Excerpted exchange} (condensed).

Auditor: Walk me through the harness end to end.

Lead: The sweep controller reads the configuration grid, spawns one sandbox per configuration, and streams metrics to the tracker. When the sweep completes, the scorer computes a composite of eval accuracy, regression-suite pass rate, and a safety-filter score, and the top checkpoint is promoted to the shared queue.

Auditor: Why a composite score rather than gating on each metric separately?

Lead: We wanted a single ranking so promotion could be automatic. Separate gates stalled too many sweeps. The composite weights were tuned against six historical sweeps.

Auditor: How does the composite handle a run that terminates early?

Lead: It would be excluded from the ranking.

Auditor: Can you show me where the exclusion happens?

Contributor (navigating the scorer): It looks like incomplete runs get a partial score. The missing metrics default to the median of the completed runs.

Auditor: So an early-terminated run can rank above completed runs on metrics it never produced. Has that happened?

Lead: I would have to check. I don't know.

\textbf{Rubric application.} On Architecture and design the team passes: descriptions were detailed, the composite-versus-gates decision was justified, and dataset and optimization questions (not excerpted) were answered correctly. On Software implementation the team fails: the promotion path is the safety-relevant path of this artifact, the answer given contradicted the code, and no attendee could establish whether an unintended promotion had occurred. Under the Appendix~\ref{app:rubrics} implementation rubric this is a significant gap in explaining how the code functions combined with a meaningfully incorrect answer, which grades as Failing regardless of the passing architecture dimension.

\textbf{Outcome.} The contribution receives a failing grade. Promotion from the harness is gated pending remediation: the team must determine whether any early-terminated run was promoted, correct the default-scoring behavior, and demonstrate understanding when remediation is verified under the Section~\ref{sec:design} lifecycle. The case illustrates two design points from the body. First, single-hop questions were insufficient. The gap surfaced on the second hop of a follow-up, which is why examinations use adaptive follow-up questioning rather than a fixed questionnaire. Second, every process metric was green while the gap accumulated, which is the asymmetry noted in Section~\ref{sec:metrics}: healthy metrics cannot certify comprehension.

\subsection{Case B: an analysis gap that grades as a pass}
\label{app:caseB}

\textbf{Setup.} The audited artifact is the evaluation report accompanying a release candidate, prepared by a three-person team. During preparation the auditors noticed in the run logs that one benchmark suite had been executed twice. The engineer responsible was required to attend by the team lead.

\textbf{Excerpted exchange} (condensed).

Auditor: Your run log shows this suite ran twice, and the first run shows a six-point swing on one reasoning benchmark. What happened?

Engineer: The first run had a variance spike concentrated in one benchmark shard. Three causes were plausible: a data-loader ordering bug, resource contention on the shared cluster, or true run-to-run variance. I re-ran with a fresh seed on a reserved node, got results within the historical band, and shipped the report.

Auditor: Which cause was it?

Engineer: I don't know. I logged it as unresolved. If it was loader ordering it would only affect shard seven, and the shard-level scores in the second run were within a point of historical values, so the reported numbers don't depend on the answer. If it recurs, the loader is the first place to look.

Auditor: Why defer rather than root-cause?

Engineer: The release gate needed the report that week. Root-causing meant reproducing the spike on the shared cluster, roughly two days. I judged that the bounded impact didn't justify the delay, my lead agreed, and the log records that decision.

Auditor: In what way does this bound the impact? Can you justify why you excluded the first run result?

Engineer: I believe the one run was an outlier but no, I can't.

Auditor: Did you record the anomaly in the bug tracking system?

Engineer: Let me check (pulls up bug tracking system and searches). No I guess I didn't.

\textbf{Rubric application.} Under Key findings and Analysis of results this grades as Passing: the anomaly was identified, plausible causes were enumerated, the team performed an (inaccurate) impact analysis, and the deferral decision was reasoned about and recorded. Note that the impact analysis was flawed: changing the random seed and execution environment does not establish the validity of the benchmark finding. However, this is a shortcoming in research process, not a knowledge gap. The investigation identified an error of scientific judgment that the engineer understands when probed.

\textbf{Outcome.} Pass. The audit report carries recommendations that the team improve its rigor in analyzing errors as well as the processes it uses to track unresolved issues to be triaged with explicit ownership, so that deferred anomalies are revisited rather than forgotten. This case illustrates how comprehension failures are narrower than other kinds of process shortcomings. An incorrect analysis performed with sufficient understanding is a diligence matter, handled through recommendations and coaching. The audit fails teams that cannot explain what they built and do not understand the analysis conducted. It does not fail teams for errors of scientific judgment they can identify and reason about. This distinction is important so there is a clean signal of losing understanding (whereas any R\&D process will always have many aspects that might be improved). It also helps improve honesty, by removing incentives to hide issues.

\section{Study Methodology}
\label{app:methodology}

The study analyzed development and review trends for open source GitHub repositories of prominent AI development organizations in the period from Q4 2025 to Q2 2026 (October 2025 to June 2026), with the period of Q1 2024 (January 2024 to March 2024) as a control before meaningful use of agentic coding tools. All data for PRs had an eight week cutoff so any reviews, comments, or whether merged was finalized eight weeks after initial creation, to make data comparable across periods. Analysis for Q1 2024 and Q4 2025 showed that more than 97\% of all reviews, comments or merges that occurred by June 2026 happened within eight weeks of creation. All data is reported as of 2026-09-01, which is more than eight weeks after the end of the study period. Data was collected using the GitHub GraphQL API incrementally with a final collection for the period on September 1 2026.

\subsection{Scope of the study}
\label{app:c-scope}

We examined open source GitHub repositories for prominent AI development organizations. We included high profile AI labs that had evidence of quality from external lists or a flagship repository with at least 1000 stars. We only included organizations where there was evidence of at least 5 PRs or 0.1\% of total PRs that included AI-written code to exclude organizations where internal tooling or policies (e.g.,\ squash merging) made AI usage not visible.

All the repos of AI-focused organizations were included in the analysis: this included organizations on curated lists and those where at least 2 of their top 5 starred repositories had maintained-declared AI topics. The repos in the orgs were classified based on whether they had maintainer-declared AI topics in a defined list. An overrides list excluded 12 organizations that AI-washed their metadata and included 3 topic-sparse AI developers. For other organizations (mostly larger organizations), only identified AI-topic repositories were included in scope for the study.

This analysis was performed on 224 GitHub organizations (orgs) and 2,135 repositories (1,278 from full-org members and 857 AI-topic repos of repo-scoped orgs). 179 of the orgs were classified as AI developers and all of their public repositories (repos) were included in scope. 45 of them were classified as a more general org that did some prominent AI R\&D, for which only specific repos that were identified as AI development were included. The latter was to exclude non-AI work from orgs like Apache, Google, Microsoft and Alibaba.

The orgs were selected using five third party lists plus a hand-curated list of 18 notable orgs (trimmed from an initial 33 to those admitted by no other rule). Orgs that created models on Epoch AI's notable models list, that were included in TechCrunch's 2025 \$100M+ AI funding round-up or that were on the hand-curated list were all considered. Those on the LinuxFoundation AI \& Data landscape, Hugging Face hub top organizations and GitHub AI-topic organizations lists needed to have at least one public repo with at least 1000 stars to be considered (to filter less prominent orgs). We used Claude Fable 5 to do mapping of company names to GitHub organizations.

\subsection{Time frames and data volume}
\label{app:c-time}

The study covers Q1 2024 as a control period and Q4 2025 through Q2 2026 as the study period. Q1 2024 predates the agentic coding tools that the detection techniques below identify; copilot-style IDE assistance and chatbot use were already present in Q1 2024 but are not detectable by this methodology.

\begin{table*}[htbp]
  \caption{Traceable AI usage over time (contributor-weighted means of per-repository shares; 95\% bootstrap confidence intervals in brackets).}
  \label{tab:t6}
  \centering
  \scriptsize\setlength{\tabcolsep}{3pt}
  \begin{tabular}{llllll}
    \toprule
    Period & Repos & AI share of submitted PRs & AI share of merged PRs & AI-PR share of submitted LOC & AI-PR share of merged LOC \\
    \midrule
    Q1 2024 & 465 & 0.0\% [0.0, 0.0] & 0.0\% [0.0, 0.0] & 0.0\% [0.0, 0.0] & 0.0\% [0.0, 0.0] \\
    \addlinespace
    Q4 2025 & 825 & 3.0\% [2.3, 3.7] & 2.2\% [1.6, 2.9] & 4.3\% [3.4, 5.4] & 2.7\% [1.9, 3.6] \\
    \addlinespace
    Q1 2026 & 1009 & 10.3\% [9.0, 11.7] & 8.8\% [7.5, 10.4] & 13.6\% [11.7, 15.4] & 11.1\% [9.5, 12.9] \\
    \addlinespace
    Q2 2026 & 1152 & 16.8\% [14.8, 18.7] & 16.4\% [14.3, 18.4] & 22.5\% [19.6, 25.0] & 20.6\% [17.9, 23.2] \\
    \bottomrule
  \end{tabular}
\end{table*}

The pooled (all-PR) AI share of submitted PRs in Q2 2026 is 17.2\%.

\begin{table*}[htbp]
  \caption{Population counts per quarter.}
  \label{tab:t7}
  \centering
  \scriptsize\setlength{\tabcolsep}{3pt}
  \begin{tabular}{llllllll}
    \toprule
    Quarter & Repos with 5+ PRs & PRs (all authors) & Merged (8-week mature) & Human-authored PRs & Human authors & Active insider authors (non-fleet) & Active fleet accounts \\
    \midrule
    Q1 2024 & 465 & 72,290 & 54,355 & 66,095 & 9,571 & 1,432 & 13 \\
    \addlinespace
    Q4 2025 & 825 & 134,088 & 96,516 & 121,775 & 16,996 & 3,002 & 27 \\
    \addlinespace
    Q1 2026 & 1,009 & 197,860 & 126,696 & 177,275 & 25,035 & 3,325 & 41 \\
    \addlinespace
    Q2 2026 & 1,152 & 285,088 & 161,565 & 255,580 & 37,897 & 3,526 & 40 \\
    \bottomrule
  \end{tabular}
\end{table*}

Coverage consists of 8,540 repo-quarters, of which 8,240 were collected and 300 were waived (due to 75 repos being inaccessible). There were 4,924 insider authors who had any insider-association PR in the study months. There were 697 in the balanced panel with at least one PR in all four quarters. There were 337 in the regular contributors panel used in Section~\ref{sec:empirical} (202 with detected AI use, 135 without).

Confidence intervals: Table~\ref{tab:t6} intervals are percentile bootstraps over repositories (2,000 seeded draws) of the contributor-weighted mean of per-repository shares. For the pooled Figure~\ref{fig:review} series, review rates carry Wilson 95\% intervals and comments per kLOC carry percentile bootstraps over PRs (2,000 draws); Tables 9 and 10 report them at the figure's monthly grain. Monthly Figure~\ref{fig:review} points are based on 15,000-35,000 non-AI, 700-9,700 AI and 700-7,800 fleet merged PRs per month.

Unit of analysis: Figure~\ref{fig:review} series are pooled over PRs per calendar month (repositories weighted by PR volume). Table~\ref{tab:t6} is contributor-weighted across repositories (weight = core committers active in Q4 2025 through Q2 2026 with 5+ commits; repo floor 5 PRs per quarter). The productivity medians in Section~\ref{sec:empirical} are per contributor with no repository weighting.

\subsection{AI detection and its limits}
\label{app:c-detect}

There are three prominent means that are used to identify the presence of AI tools in writing code\footnote{This section focuses on metrics for projects using git since leading AI tools integrate natively with git and provide usage metadata only for it.}:

\begin{itemize}
  \item Metadata that is included in code commits (such as commit comment footers like ``Generated with Claude Code'' or co-author trailers like cursoragent@cursor.com)
  \item Pull Requests that are authored by coding-agent bot accounts (such as Copilot coding agent or Devin)
  \item Branches that have a known prefix. Some AI tools hosted in Web or cloud create branches with standard prefixes such as codex/ or copilot/ (this is how PRarena \citep{Avetisian_2025} tracks agentic tools). This has more possibility for false positives.
\end{itemize}

The presence of the first two indications is a lower bound on AI usage. There are workflows that use AI to simply generate code that will not produce them (e.g.,\ for inline edits and code suggestions or even to write changes locally that are committed by the human developer). Some tools (notably OpenAI Codex CLI) do not emit trackable metadata by default, even if they are used to create PRs. The use of squash merges in Git also erases metadata that would otherwise allow tracking AI usage.

The detection methods are floors on AI usage since they do not detect inline-edits, use of untracked tools or squash merges.

The detection layers and the signatures matched were:

\begin{itemize}
  \item Layer A, commit markers (co-author emails, footers or committer identities in the PR's first 12 commits): Claude (noreply@anthropic.com, ``Generated with [Claude Code]''); Cursor (cursoragent@cursor.com, agent@cursor.com); aider (Co-authored-by: aider); Codex (codex@openai.com, committer chatgpt-codex-connector[bot]); OpenHands (openhands@all-hands.dev, committer openhands-agent); and a vetted list of self-identified agent identities (hermes-agent@*, agent@hermes, agent@modal.com, agent@deepagents.dev, pragma-agent@modular.com, agent@qwenpaw.local, agent@example.com, bot@imbue.com).
  \item Layer B, agent PR authors (Bot-typed accounts): copilot-swe-agent, devin-ai-integration, cursor, google-labs-jules, codegen-sh, amazon-q-developer, factory-droid, claude.
  \item Layer C, head-branch prefixes: codex/, copilot/, cursor/, claude/, droid/. The last three are treated as weak; a strict variant additionally requires a GitHub web-flow committer or the product's generated slug.
\end{itemize}

Precision: measured for Layer C only, on org-stratified samples of PRs detected by Layer C alone. The share with positive provenance was codex/ 95\%, copilot/ 98\%, cursor/ 49\% and claude/ 42\%; the cursor/ and claude/ remainder is indeterminate (a human-named branch is indistinguishable from an agent branch continued locally). One false-positive class, prefix branches committed entirely by github-actions[bot] (18.4\% of Layer-C-only claude/ samples), is excluded. The Q2 2026 headline is therefore a two-sided bound: 16.83\% with weak prefixes counted and 16.72\% with positive provenance required (0.11 percentage points wide; at most 0.01 points in earlier quarters). Zero prefix-named PRs appear in Q1 2024. Layers A and B rest on exact-identity matches audited by hand but were not sampled for false positives; human-copied trailers count as AI involvement by design.

\subsection{Contributor and reviewer classification}
\label{app:c-class}

Contributors were classified as insiders based on being a member, owner or collaborator in a repo. An author is an insider if any study-period PR carries that association. Staff whose organization membership is private appear as CONTRIBUTOR and are missed. Regular contributors are those who submitted at least five PRs in each of the four quarters and a mean of at least ten PRs per quarter including the baseline quarter, excluding fleet operators.

Fleet operators: 42 accounts merged 300 or more PRs in at least one study quarter; 33 of the 42 also hold member or collaborator status in a study repository. The flag is global: an account over the threshold in any quarter is excluded from every population in every quarter. Fleet operators are defined by submission volume alone; the content of their PRs was not classified and no inference about their use of AI is made. Fleet PRs were 4.7\% of merged eligible PRs in Q1 2024 (2,367 of 50,406) and 15.2\% in Q2 2026 (21,992 of 145,044).

Table~\ref{tab:t8} analyzes the sensitivity of the productivity increase to the panel rule:

\begin{table*}[htbp]
  \caption{Sensitivity of the median submitted kLOC per quarter to the regular contributor panel rule (Q1 2024 to Q2 2026).}
  \label{tab:t8}
  \centering
  \scriptsize\setlength{\tabcolsep}{3pt}
  \begin{tabular}{lllllll}
    \toprule
    Rule (min every quarter / mean over four) & n & Ever-AI / never-AI & Submitted kLOC Q1 2024 to Q2 2026 & Submitted multiple & Merged multiple & Ever-AI / never-AI submitted multiple \\
    \midrule
    3 / 5 & 438 & 250 / 188 & 4.69 to 13.00 & 2.77 & 2.75 & 3.54 / 2.32 \\
    \addlinespace
    5 / 10 (paper) & 337 & 202 / 135 & 6.26 to 17.72 & 2.83 & 2.73 & 3.02 / 2.32 \\
    \addlinespace
    10 / 20 & 211 & 140 / 71 & 8.89 to 26.05 & 2.93 & 2.28 & 3.09 / 2.40 \\
    \addlinespace
    5 / 5 & 348 & 204 / 144 & 6.12 to 16.80 & 2.75 & 2.67 & 3.13 / 2.40 \\
    \addlinespace
    10 / 10 & 233 & 150 / 83 & 8.66 to 23.58 & 2.72 & 2.40 & 3.22 / 2.27 \\
    \bottomrule
  \end{tabular}
\end{table*}

Bootstrap 95\% intervals for the paper rule: submitted multiple 2.83 [2.10, 3.64] overall and 3.02 [2.47, 4.36] for contributors with detected AI use; merged multiples 2.73 and 3.25.

Reviewer classification: self-reviews are excluded by design. Reviews from accounts tagged as bot accounts are not counted as human reviews. Reviews from User-typed accounts whose logins end in -bot, \_bot, [bot], -ci, -robot, -agent or \_agent, or begin with bot-, are likewise excluded from human review (e.g.,\ PaddlePaddle-bot, all-hands-bot, qwen-code-ci-bot); the match is anchored so that logins such as talbot remain human. Reviews from known AI reviewer accounts are tracked separately as AI reviews.

\subsection{Review metric definitions}
\label{app:c-metrics}

A merged PR is counted as human-reviewed if it received at least one review from a human (non-bot) account other than the author within eight weeks of creation. Any PRs merged by the cutoff are counted in the denominator, including self-merged PRs. Comments per kLOC counts inline review comments by human reviewers other than the author per thousand lines of code in the PR for PRs that were merged and human-reviewed, winsorized at 20 comments per PR per reviewer to limit the influence of unusually long review threads. Review summary text is not counted. Inline commentary share is the proportion of merged PRs with at least one inline human comment. LOC is the sum of line additions plus deletions for the PR, capped at 20,000 lines per PR, with no file exclusions (generated and vendored files count). Quarters are assigned by PR creation date, so a PR created in March and merged in April is a Q1 PR.

\subsection{Figure~\ref{fig:review} under alternative weightings and on a constant repository set}
\label{app:c-weights}

The following tables analyze alternative weightings for human review rates and human comment rates.

\begin{table*}[p]
  \caption{Review rate for merged PRs by month and slice.}
  \label{tab:t9}
  \centering
  \scriptsize\setlength{\tabcolsep}{3pt}
  \begin{tabular}{lllllll}
    \toprule
    Month & Slice & Merged PRs & Pooled \% & Wilson 95\% & Contributor-weighted repo mean \% & Repos \\
    \midrule
    Jan 2024 & Non-AI & 15,451 & 77.2 & [76.6, 77.9] & 70.0 [61.7, 77.5] & 366 \\
    Jan 2024 & Fleet & 745 & 44.2 & [40.6, 47.7] & 44.9 [21.0, 80.3] & 17 \\
    Feb 2024 & Non-AI & 15,171 & 77.3 & [76.6, 77.9] & 72.0 [63.4, 79.8] & 378 \\
    Feb 2024 & Fleet & 877 & 54.6 & [51.3, 57.9] & 37.1 [11.2, 79.5] & 24 \\
    Mar 2024 & Non-AI & 17,417 & 78.0 & [77.4, 78.6] & 74.1 [66.4, 80.9] & 390 \\
    Mar 2024 & Fleet & 745 & 48.2 & [44.6, 51.8] & 42.7 [20.2, 75.4] & 19 \\
    Oct 2025 & Non-AI & 28,743 & 82.1 & [81.7, 82.6] & 78.3 [73.3, 82.8] & 625 \\
    Oct 2025 & AI & 863 & 71.4 & [68.3, 74.3] & 78.4 [68.9, 86.1] & 65 \\
    Oct 2025 & Fleet & 1,617 & 42.6 & [40.2, 45.0] & 50.9 [37.2, 67.1] & 53 \\
    Nov 2025 & Non-AI & 26,304 & 80.8 & [80.4, 81.3] & 77.6 [72.2, 82.3] & 625 \\
    Nov 2025 & AI & 738 & 66.5 & [63.0, 69.8] & 77.4 [68.4, 84.5] & 79 \\
    Nov 2025 & Fleet & 1,803 & 43.9 & [41.6, 46.2] & 49.9 [35.4, 67.2] & 55 \\
    Dec 2025 & Non-AI & 26,455 & 80.0 & [79.5, 80.5] & 79.5 [74.9, 83.8] & 633 \\
    Dec 2025 & AI & 1,206 & 78.6 & [76.2, 80.8] & 74.2 [64.2, 84.2] & 86 \\
    Dec 2025 & Fleet & 1,938 & 33.5 & [31.4, 35.6] & 48.6 [33.8, 66.5] & 51 \\
    Jan 2026 & Non-AI & 28,788 & 79.8 & [79.4, 80.3] & 77.6 [72.6, 82.2] & 651 \\
    Jan 2026 & AI & 1,946 & 75.0 & [73.0, 76.8] & 78.0 [69.8, 84.3] & 143 \\
    Jan 2026 & Fleet & 2,491 & 31.3 & [29.5, 33.1] & 47.0 [29.8, 66.8] & 50 \\
    Feb 2026 & Non-AI & 28,571 & 79.8 & [79.3, 80.3] & 76.3 [71.4, 80.4] & 701 \\
    Feb 2026 & AI & 4,247 & 61.9 & [60.4, 63.3] & 74.6 [67.7, 81.2] & 306 \\
    Feb 2026 & Fleet & 3,149 & 28.7 & [27.2, 30.3] & 40.3 [26.0, 58.6] & 56 \\
    Mar 2026 & Non-AI & 35,446 & 79.1 & [78.6, 79.5] & 76.2 [71.4, 80.1] & 748 \\
    Mar 2026 & AI & 5,800 & 72.4 & [71.2, 73.5] & 77.1 [70.4, 82.8] & 344 \\
    Mar 2026 & Fleet & 5,760 & 24.1 & [23.0, 25.2] & 40.9 [25.4, 60.7] & 68 \\
    Apr 2026 & Non-AI & 33,683 & 78.5 & [78.0, 78.9] & 77.5 [72.6, 81.2] & 730 \\
    Apr 2026 & AI & 6,796 & 70.4 & [69.3, 71.5] & 78.0 [71.8, 83.3] & 387 \\
    Apr 2026 & Fleet & 6,560 & 26.8 & [25.8, 27.9] & 47.2 [33.1, 64.3] & 71 \\
    May 2026 & Non-AI & 31,899 & 78.4 & [77.9, 78.8] & 76.2 [72.1, 79.8] & 735 \\
    May 2026 & AI & 7,442 & 69.6 & [68.6, 70.7] & 78.4 [73.6, 82.3] & 415 \\
    May 2026 & Fleet & 7,651 & 21.1 & [20.2, 22.0] & 43.0 [27.7, 61.8] & 66 \\
    Jun 2026 & Non-AI & 33,535 & 78.8 & [78.3, 79.2] & 78.2 [74.6, 81.5] & 730 \\
    Jun 2026 & AI & 9,697 & 71.2 & [70.3, 72.1] & 78.1 [73.4, 82.1] & 459 \\
    Jun 2026 & Fleet & 7,781 & 19.8 & [18.9, 20.7] & 43.9 [26.8, 63.0] & 65 \\
    \bottomrule
  \end{tabular}
\end{table*}

Table~\ref{tab:t9} shows the pooled review rates plotted in Figure~\ref{fig:review}. They are pooled over PRs with Wilson 95\% intervals. It shows the contributor-weighted repository mean with bootstrap 95\% intervals, using repos with two or more merged PRs in the slice-month.

\begin{table*}[p]
  \caption{Human inline comments per kLOC on merged, human-reviewed PRs by month and slice.}
  \label{tab:t10}
  \centering
  \scriptsize\setlength{\tabcolsep}{3pt}
  \begin{tabular}{lllllll}
    \toprule
    Month & Slice & Reviewed PRs & kLOC & Comments per kLOC & Bootstrap 95\% & \% of Q1 2024 non-AI \\
    \midrule
    Jan 2024 & Non-AI & 11,906 & 4,137 & 4.13 & [3.81, 4.49] & 104\% \\
    Jan 2024 & Fleet & 328 & 106 & 1.92 & [1.24, 2.98] & 48\% \\
    Feb 2024 & Non-AI & 11,702 & 4,436 & 3.88 & [3.56, 4.20] & 98\% \\
    Feb 2024 & Fleet & 479 & 78 & 2.23 & [1.26, 3.95] & 56\% \\
    Mar 2024 & Non-AI & 13,557 & 5,030 & 3.90 & [3.61, 4.23] & 98\% \\
    Mar 2024 & Fleet & 359 & 82 & 1.35 & [0.79, 2.33] & 34\% \\
    Oct 2025 & Non-AI & 23,550 & 10,971 & 2.86 & [2.73, 3.02] & 72\% \\
    Oct 2025 & AI & 616 & 303 & 2.19 & [1.66, 2.91] & 55\% \\
    Oct 2025 & Fleet & 689 & 343 & 0.75 & [0.51, 1.06] & 19\% \\
    Nov 2025 & Non-AI & 21,227 & 10,062 & 2.77 & [2.63, 2.92] & 70\% \\
    Nov 2025 & AI & 491 & 295 & 2.12 & [1.49, 3.04] & 53\% \\
    Nov 2025 & Fleet & 791 & 440 & 0.75 & [0.52, 1.09] & 19\% \\
    Dec 2025 & Non-AI & 21,129 & 10,758 & 2.45 & [2.33, 2.59] & 62\% \\
    Dec 2025 & AI & 947 & 497 & 1.61 & [1.23, 2.16] & 41\% \\
    Dec 2025 & Fleet & 649 & 322 & 0.87 & [0.63, 1.20] & 22\% \\
    Jan 2026 & Non-AI & 22,951 & 12,723 & 2.25 & [2.14, 2.36] & 57\% \\
    Jan 2026 & AI & 1,456 & 1,074 & 1.56 & [1.31, 1.87] & 39\% \\
    Jan 2026 & Fleet & 776 & 427 & 0.77 & [0.54, 1.09] & 19\% \\
    Feb 2026 & Non-AI & 22,766 & 13,676 & 2.09 & [1.99, 2.21] & 53\% \\
    Feb 2026 & AI & 2,624 & 1,852 & 1.84 & [1.61, 2.11] & 47\% \\
    Feb 2026 & Fleet & 903 & 379 & 0.79 & [0.57, 1.11] & 20\% \\
    Mar 2026 & Non-AI & 27,989 & 16,350 & 1.96 & [1.88, 2.05] & 50\% \\
    Mar 2026 & AI & 4,196 & 2,638 & 1.75 & [1.56, 1.97] & 44\% \\
    Mar 2026 & Fleet & 1,384 & 1,024 & 0.46 & [0.35, 0.61] & 12\% \\
    Apr 2026 & Non-AI & 26,397 & 14,813 & 1.97 & [1.88, 2.06] & 50\% \\
    Apr 2026 & AI & 4,783 & 3,331 & 1.75 & [1.58, 1.94] & 44\% \\
    Apr 2026 & Fleet & 1,755 & 934 & 0.54 & [0.42, 0.70] & 14\% \\
    May 2026 & Non-AI & 24,966 & 15,704 & 1.89 & [1.81, 1.98] & 48\% \\
    May 2026 & AI & 5,180 & 4,121 & 1.54 & [1.41, 1.70] & 39\% \\
    May 2026 & Fleet & 1,608 & 1,210 & 0.55 & [0.43, 0.68] & 14\% \\
    Jun 2026 & Non-AI & 26,382 & 15,919 & 1.82 & [1.74, 1.90] & 46\% \\
    Jun 2026 & AI & 6,893 & 5,324 & 1.44 & [1.32, 1.57] & 36\% \\
    Jun 2026 & Fleet & 1,539 & 1,126 & 0.80 & [0.65, 0.99] & 20\% \\
    \bottomrule
  \end{tabular}
\end{table*}

Table~\ref{tab:t10} shows the pooled comment rates plotted in Figure~\ref{fig:review}. They are pooled over PRs with Wilson 95\% intervals. It shows the contributor-weighted repository mean . The results show a percentile bootstrap over PRs with 2,000 draws. The last column is relative to the Q1 2024 non-AI pooled rate of 3.96.

\begin{table*}[p]
  \caption{Figure~\ref{fig:review} quantities by month recomputed on active repositories.}
  \label{tab:t11}
  \centering
  \scriptsize\setlength{\tabcolsep}{3pt}
  \begin{tabular}{lllllll}
    \toprule
    Month & Slice & Merged PRs & Review \% [Wilson 95\%] & Contributor-weighted repo mean \% & Comments per kLOC [bootstrap 95\%] & \% of Q1 2024 non-AI \\
    \midrule
    Jan 2024 & Non-AI & 14,243 & 77.7 [77.0, 78.4] & 70.1 & 4.26 [3.92, 4.63] & 105\% \\
    Jan 2024 & Fleet & 745 & 44.2 [40.6, 47.7] & 44.9 & 1.92 [1.24, 2.98] & 47\% \\
    Feb 2024 & Non-AI & 13,947 & 77.4 [76.7, 78.1] & 72.1 & 3.95 [3.64, 4.29] & 98\% \\
    Feb 2024 & Fleet & 872 & 54.6 [51.3, 57.9] & 36.8 & 2.21 [1.24, 4.02] & 55\% \\
    Mar 2024 & Non-AI & 16,112 & 78.2 [77.6, 78.8] & 74.3 & 3.95 [3.65, 4.28] & 98\% \\
    Mar 2024 & Fleet & 731 & 48.6 [45.0, 52.2] & 42.7 & 1.30 [0.73, 2.21] & 32\% \\
    Oct 2025 & Non-AI & 18,549 & 83.9 [83.3, 84.4] & 83.3 & 3.16 [2.97, 3.38] & 78\% \\
    Oct 2025 & AI & 554 & 78.9 [75.3, 82.1] & 79.6 & 2.56 [1.82, 3.66] & 63\% \\
    Oct 2025 & Fleet & 1,074 & 40.9 [38.0, 43.8] & 46.8 & 0.52 [0.33, 0.82] & 13\% \\
    Nov 2025 & Non-AI & 17,053 & 82.6 [82.0, 83.1] & 83.1 & 2.97 [2.78, 3.17] & 73\% \\
    Nov 2025 & AI & 509 & 72.7 [68.7, 76.4] & 77.6 & 2.24 [1.55, 3.35] & 56\% \\
    Nov 2025 & Fleet & 1,257 & 44.2 [41.4, 46.9] & 46.4 & 0.62 [0.39, 1.03] & 15\% \\
    Dec 2025 & Non-AI & 16,701 & 81.7 [81.1, 82.3] & 84.0 & 2.52 [2.36, 2.70] & 62\% \\
    Dec 2025 & AI & 891 & 85.5 [83.1, 87.7] & 74.7 & 1.64 [1.21, 2.21] & 41\% \\
    Dec 2025 & Fleet & 1,342 & 34.1 [31.6, 36.6] & 43.6 & 0.69 [0.46, 1.00] & 17\% \\
    Jan 2026 & Non-AI & 17,421 & 81.5 [80.9, 82.1] & 82.3 & 2.37 [2.21, 2.53] & 59\% \\
    Jan 2026 & AI & 1,362 & 79.9 [77.7, 81.9] & 77.9 & 1.71 [1.37, 2.11] & 42\% \\
    Jan 2026 & Fleet & 1,493 & 31.1 [28.8, 33.5] & 42.6 & 0.44 [0.28, 0.70] & 11\% \\
    Feb 2026 & Non-AI & 16,315 & 83.8 [83.3, 84.4] & 82.5 & 2.40 [2.25, 2.57] & 59\% \\
    Feb 2026 & AI & 2,551 & 71.3 [69.5, 73.0] & 79.2 & 2.05 [1.73, 2.43] & 51\% \\
    Feb 2026 & Fleet & 1,503 & 29.3 [27.1, 31.7] & 43.8 & 0.81 [0.50, 1.43] & 20\% \\
    Mar 2026 & Non-AI & 19,331 & 83.3 [82.8, 83.8] & 81.3 & 2.39 [2.24, 2.53] & 59\% \\
    Mar 2026 & AI & 3,434 & 80.0 [78.7, 81.3] & 82.2 & 1.99 [1.73, 2.29] & 49\% \\
    Mar 2026 & Fleet & 1,625 & 39.7 [37.3, 42.1] & 44.7 & 0.46 [0.31, 0.67] & 11\% \\
    Apr 2026 & Non-AI & 17,979 & 83.7 [83.1, 84.2] & 83.4 & 2.24 [2.10, 2.39] & 56\% \\
    Apr 2026 & AI & 3,547 & 76.8 [75.4, 78.2] & 82.0 & 2.04 [1.78, 2.33] & 50\% \\
    Apr 2026 & Fleet & 1,843 & 53.3 [51.1, 55.6] & 51.0 & 0.40 [0.29, 0.55] & 10\% \\
    May 2026 & Non-AI & 16,486 & 81.4 [80.8, 82.0] & 82.4 & 2.23 [2.10, 2.38] & 55\% \\
    May 2026 & AI & 3,795 & 73.6 [72.1, 74.9] & 81.9 & 1.63 [1.43, 1.85] & 40\% \\
    May 2026 & Fleet & 1,905 & 42.2 [40.0, 44.4] & 45.3 & 0.37 [0.26, 0.53] & 9\% \\
    Jun 2026 & Non-AI & 17,599 & 82.4 [81.8, 83.0] & 83.1 & 2.11 [1.97, 2.24] & 52\% \\
    Jun 2026 & AI & 4,366 & 75.2 [73.9, 76.5] & 82.0 & 1.45 [1.29, 1.64] & 36\% \\
    Jun 2026 & Fleet & 1,677 & 43.6 [41.2, 46.0] & 50.0 & 0.38 [0.27, 0.53] & 9\% \\
    \bottomrule
  \end{tabular}
\end{table*}

Table~\ref{tab:t11} analyzes human review rates and human comment rates for the 332 repos with at least five PRs in every study quarter. These review and comment rates are pooled. The contributor-weighted repository mean uses repos with two or more merged PRs in the slice-month. The last column is relative to the constant-set Q1 2024 non-AI pooled rate of 4.04. For this more active set of repos the AI review rates were 3-12 points lower with a median of 8 points. However for the contributor weighting, there is no consistent gap between non-AI and AI review rates. These repos also show a similar decline in comments per LOC as the other slices. On the constant repository set the AI share of submitted PRs in Q2 2026 is 15.6\% (16.8\% in the full cross-section).

\end{document}